\documentclass[aps,pra,10pt,twocolumn,floatfix]{revtex4-1}

\usepackage{times}
\usepackage{graphicx}
\usepackage{subfigure}
\usepackage{dcolumn}
\usepackage{bm}
\usepackage{xcolor}
\usepackage{amsmath,amssymb}
\usepackage{verbatim}
\usepackage{mathtools}
\usepackage{epstopdf}
\usepackage{bbold}

\usepackage{hyperref}
\hypersetup{
	colorlinks=true,
	linkcolor=blue,           
	citecolor=blue,           
	filecolor=magenta,        
	urlcolor=cyan
}

\renewcommand{\i}{\mathrm{i}}

\DeclarePairedDelimiterX\braket[2]{\langle}{\rangle}{#1 \delimsize\vert #2}
\DeclarePairedDelimiterX\ketbra[2]{\lvert}{\rvert}{#1 \rangle\hspace{-.25em}\langle #2}

\begin{document}
	
	\title{Training Coupled Phase Oscillators as a Neuromorphic Platform using Equilibrium Propagation}
	
	\author{Qingshan Wang}
	\affiliation{Max Planck Institute for the Science of Light, Staudtstraße 2, 91058 Erlangen, Germany}
	
	\author{Clara C. Wanjura}
	\affiliation{Max Planck Institute for the Science of Light, Staudtstraße 2, 91058 Erlangen, Germany}
	 
	\author{Florian Marquardt}
	\affiliation{Max Planck Institute for the Science of Light, Staudtstraße 2, 91058 Erlangen, Germany}
    \affiliation{Department of Physics, University of Erlangen-Nuremberg, 91058 Erlangen, Germany}
	
	\date{\today}
	
	\begin{abstract}
        Given the rapidly growing scale and resource requirements of machine learning applications, the idea of building more efficient learning machines much closer to the laws of physics is an attractive proposition. One central question for identifying promising candidates for such neuromorphic platforms is whether not only inference but also training can exploit the physical dynamics. In this work, we show that it is possible to successfully train a system of coupled phase oscillators---one of the most widely investigated nonlinear dynamical systems with a multitude of physical implementations, comprising laser arrays, coupled mechanical limit cycles,  superfluids, and exciton-polaritons. To this end, we apply the approach of equilibrium propagation, which permits to extract training gradients via a physical realization of backpropagation, based only on local interactions. The complex energy landscape of the XY/ Kuramoto model leads to multistability, and we show how to address this challenge. Our study identifies coupled phase oscillators as a new general-purpose neuromorphic platform and opens the door towards future experimental implementations.
        
	\end{abstract}
	
	
	\maketitle
	
	\section{Introduction}

We are witnessing the widespread adoption of machine learning in all branches of science and technology. At the same time, it becomes clear that the energy costs for both training and inference are exploding in an unsustainable way. This is especially apparent for the most powerful recent innovations like large-language models and diffusion models for image generation, which contain billions of parameters. Even the impressive advances in deploying specialized hardware like graphical processing units or tensor processing units are not able to keep up with the rapidly rising resource requirements. This makes it all the more urgent to explore alternative means for computation in this domain. The hope that such alternatives exist rests on several observations. First, machine learning is inherently stochastic and it therefore seems wasteful to run implementations on digital hardware that was originally conceived for implementing mathematical algorithms where precise answers matter. In addition, one can exploit the specific structure of machine learning problems to design new hardware platforms that are no longer general purpose. In particular, one can overcome the inefficient separation between processing and memory, the so-called von-Neumann bottleneck. Third, many physical platforms naturally allow for highly parallel processing.

These are the motivations for the rapidly growing field of neuromorphic computing \Citep{markovic2020physics,christensen20222022}, which explores a large variety of different physical platforms, with the aim of designing more energy-efficient  parallel processing devices, operating much closer to the underlying physics than a typical digital processor. Neuromorphic implementations do include alternative digital electronic platforms that overcome the von-Neumann bottleneck, but there is even more fundamental research to be done in the domain of analog, more directly physics-based approaches. These range from solid-state physics (building on components like memristors \Citep{prezioso2015training},  Josephson junctions \Citep{schneider2018tutorial,shainline2017superconducting}, or spin oscillators \Citep{TorrejonGrollier2017}) and optics \Citep{wagner1987multilayer,shen2017deep, bueno2018reinforcement, feldmann2019all, feldmann2021parallel, pai2023experimentally} (free space or integrated photonics) to other domains, even including soft-matter physics (e.g.  \Citep{pashine2021local,falk2023learning,stern2023learning,altman2023experimental}). 

	\begin{figure}[h!]
	    \centering
	    \includegraphics[width = 0.42\textwidth]{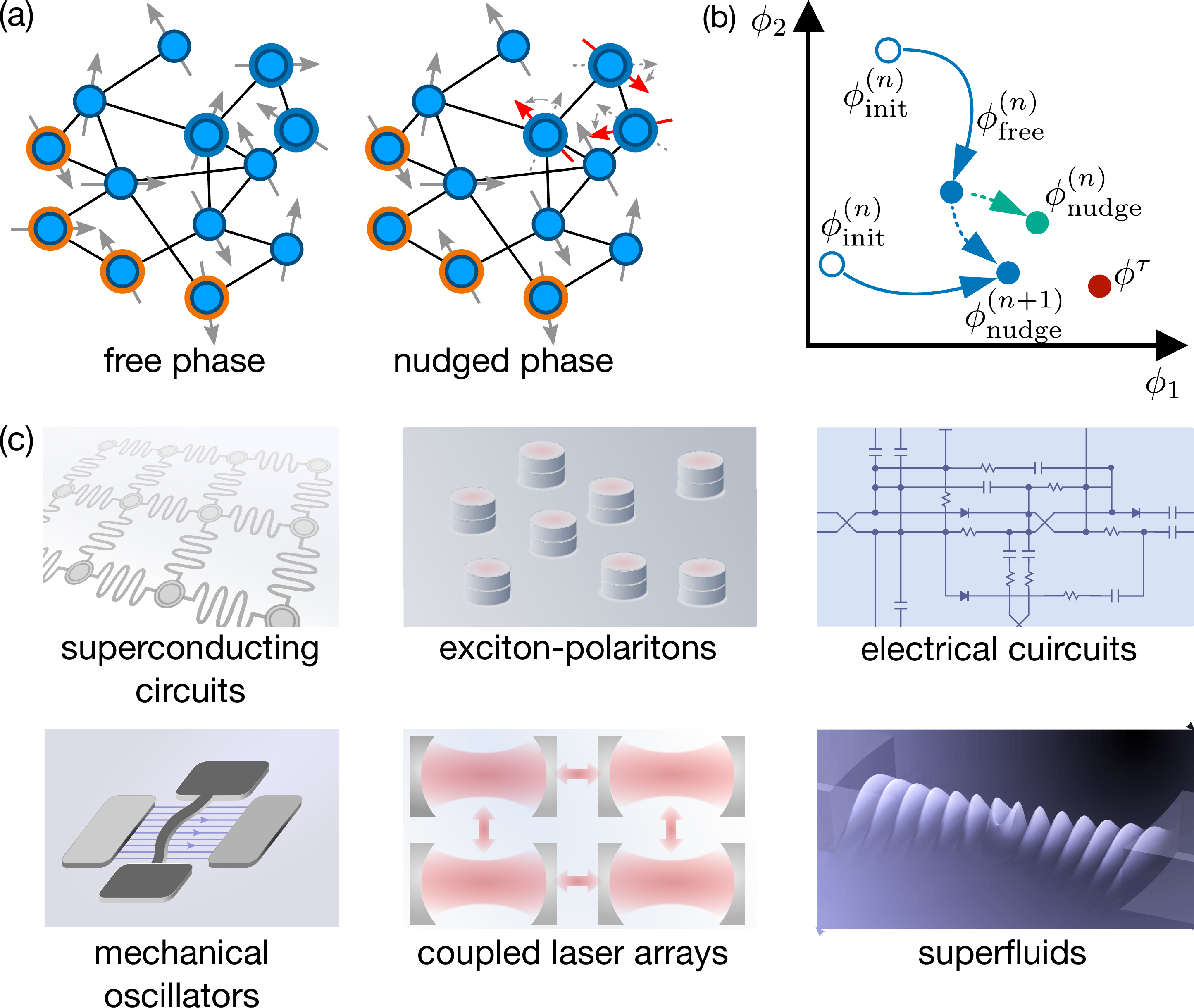}
	    \caption{\textbf{Equilibrium propagation for coupled phase oscillators.}
	    (a)~We train a neuromorphic system of coupled phase oscillators (units), here represented by blue circles.
	    In the free phase (left), the input units (orange) are fixed and the system relaxes to its equilibrium via the standard overdamped dynamics of coupled phase oscillators. Output units are highlighted by a blue rim.
	    In the nudge phase (right), the output units are forced weakly towards the desired target output (red arrows) and the the system evolves to a new, slightly shifted equilibrium. The difference between these two equilibria allows to infer the gradients needed to train the system, according to the fundamental rule of equilibrium propagation, see Eq.~\eqref{eq:EPgradient}.
        (b)~Schematic representation of the evolution in the space of oscillator phases $\phi_j$, with open circles depicting initial conditions and full circles depicting equilibria ($\phi^{\tau}$ is the desired target configuration). 
	    (c)~Possible experimental platforms that can give rise to the dynamics of the XY/ Kuramoto model of coupled phase oscillators include superconducting circuits, nonlinear electrical circuits, exciton-polariton systems, mechanical-oscillator limit cycles, coupled laser arrays and superfluids.
	    }
	    \label{fig:schematic}
	\end{figure}%

A neuromorphic platform usually has to exhibit nonlinear dynamics to be powerful enough in terms of its expressivity, i.e., its ability to approximate arbitrary input-output relations. Arguably one of the strongest generic classical nonlinearities is found in phase oscillators. These are systems where a combination of energy pump and nonlinearity establishes a limit cycle, with the phase of the oscillation as the resulting continuous degree of freedom. The laser is the most prominent example. When several such phase oscillators are coupled, they exhibit the dynamics of phase locking and synchronization. Theoretically, these systems are described by the so-called Kuramoto model \Citep{KuramotoReview_Acebron}, which is equivalent to the relaxation dynamics of an XY model \Citep{kosterlitz1974critical}, one of the most widely studied models in statistical physics.  Coupled phase oscillators have been explored in a variety of platforms, including laser arrays \Citep{nixon2013observing,takeda2017boltzmann}, coupled mechanical limit cycles in optomechanical \Citep{Heinrich2011Collective,zhang2015synchronization}  and nano-electromechanical \Citep{matheny2014phase}  systems, cold atoms \Citep{struck2013engineering}, Josephson arrays in superconducting circuits \Citep{Cosmic2020Probing}, spin oscillators \Citep{Torrejon2017Neuromorphic,romera2018vowel}, and exciton-polaritons \Citep{baas2008synchronized,Berloff2017Realizing,Kavokin2022Polariton}. 

In this article, we explore how a system of coupled phase oscillators can serve as a neuromorphic platform. Most importantly, we will investigate how it can be trained in a physics-based approach. 

An array of coupled phase oscillators was first considered as a neuromorphic platform already in \Citep{hoppensteadt1999oscillatory,hoppensteadt2000synchronization}. In that case, the system was proposed as an associative Hopfield memory, where couplings can be determined in a single step based on a simple Hebbian rule. In contrast, in our case, we will be interested in the more general setting of supervised learning, where the goal is to learn a prescribed input-output relation, for regression or classification tasks – the most wide-spread application in machine learning. That such a goal might be achievable is indicated by the results in \Citep{stroev2021neural}, where certain small architectures of coupled phase oscillators were handcrafted to address simple tasks. More generally, coupled oscillators have been proposed in various forms as computing platforms (see the review in \Citep{csaba2020coupled}), and the training of coupled digital oscillators has also been investigated recently in simulations \Citep{rudner2023design}. From the neuromorphic perspective, oscillators seem like a natural choice for machine learning, since the brain is based on spiking neurons that can also show behaviour such as synchronization.

One particular challenge in neuromorphic physical platforms is the question of efficient training. The conceptually easiest, general approach that can be adopted in every case is the parameter shift method, where internal trainable parameters of the physical device are adjusted in a feedback loop based on the distance to the desired output (e.g. \Citep{Filipovich2022Silicon,bandyopadhyay2022single} and many others). However, this scales very unfavourably, requiring for a single training step a number of evaluations rising linearly in the number of parameters. Another popular approach is reservoir computing \Citep{duport2012all,tanaka2019recent,nakajima2020physical,van2017advances}, where the fixed (non-adjustable) nonlinear dynamics of a system is used to map the input to a higher-dimensional space which is then turned into the output via a trainable simple (possibly even linear) neural network, similar to kernel methods in machine learning. This has been very successful, but still relies on a digital neural network. One major goal in the field is to find ways to impliment variants of backpropagation for neuromorphic devices, since this efficient method to find the gradients needed for training is the cornerstone of virtually all machine learning. One possibility consists in doing this via simulations (``digital twins''). This method suffers from the lower speed of simulations and the potential inaccuracy of the adopted model, although the latter constraint has recently been overcome via an ingenious hybrid method ~\Citep{wright2022deep}, where backpropagation happens in the simulation but the forward pass is performed on the real device. Physical implementations of backpropagation, exploiting the physical dynamics itself, are a conceptually very attractive goal. They have been occasionally put forward starting even in the 80s. The first ideas involved approaches that were not general purpose but relied on specific properties of specially engineered platforms, especially in optics \Citep{wagner1987multilayer,psaltis1990holography,guo2021backpropagation,spall2023training}. A more general type of physical backpropagation for an integrated-photonics platform was introduced and eventually implemented recently in  \Citep{hughes2018training,pai2023experimentally}.  In certain cases, gradients can also be extracted directly from scattering experiments, as shown for the recently introduced nonlinear neuromorphic processing in linear wave scattering platforms \Citep{wanjura2023fully}. However, only two generic methods exist for physically performing backpropagation in a wide class of systems. One of them is Hamiltonian Echo Backpropagation, which relies on the Hamiltonian dynamics of arbitrary time-reversal-invariant systems \cite{lopez2021self}. The other is Equilibrium Propagation \Citep{scellier2017equilibrium}, which encompasses arbitrary energy-based systems that relax to some equilibrium. Since coupled phase oscillators display relaxation-type phase dynamics, in this work we will employ Equilibrium Propagation.

We start by explaining how Equilibrium Propagation (EP) can be applied to systems of coupled phase oscillators. We then explore in numerical experiments the EP-based training of networks of oscillators in two illustrative examples (XOR and handwritten-digit recognition).
We specifically point out the effects of multistability on training and how to account for them.
Our results establish coupled phase oscillators as a viable neuromorphic platform for supervised training,
although the choice of network architecture plays an important role.
    
	\section{The XY model and Equilibrium Propagation}\label{section:modelandtheory}

    In this section, we first describe our model and then briefly review the theory of EP and how to apply it in our setting. We address the problem of multistability, which arises especially in the very nonconvex energy landscape of coupled oscillators.
 
\subsection{XY model of coupled phase oscillators}
       
The dynamics of coupled phase oscillators can be described by some variant of the Kuramoto model, and this can be understood (in many cases) as overdamped relaxation dynamics in the energy landscape of an XY model.
An XY model consists of interacting classical 2D spins (in our case, the phase oscillators), which are described by a set of phase angles $\phi_i$. Anticipating the neuromorphic application, we refer to the coupling between the spins as \textbf{weights} and assume that all weights are trainable.
We adopt the following energy function
\begin{equation}\label{eq:XY_network_energy}
    E(\phi ; W, h, \psi ) = - \frac{1}{2} \sum_{ij} W_{ij} \cos(\phi_i-\phi_j) - \sum_i h_i \cos(\phi_i-\psi_i).
\end{equation}
The first term describes coupling between arbitrary oscillators, with coupling constants (weights) $W_{ij}$. The second term describes a tendency towards phase locking to some external oscillatory drives, where $h_i$ is set by the amplitude of the respective drive and $\psi_i$ is the phase of the drive. In the machine-learning context, these could be understood in analogy to the bias terms of a neural network. Beyond that, we remark that, in the synchronization context, one would normally expect to have another set of terms, of the type $\Omega_i \phi_i$. This would denote the fact that different oscillators are running at different intrinsic frequencies. When these terms are large, a transition towards desynchronization will take place. However, for the neuromorphic application we are interested in, we find that such terms – and the associated tendency towards desynchronization – are not helpful, as they may lead to the absence of stable fixed points. We therefore envisage a situation where all the oscillators' intrinsic frequencies are the same. Small deviations from this ideal scenario can be tolerated, since the coupling terms would help to generate stable fixed points.

In this article, we focus on the deterministic dynamics of the system. We obtain the equilibrium by solving the following ODE, assuming that the phase angles are driven by the negative energy gradient
\begin{equation}\label{eq:EoM}
    \begin{split}
        \dot{\phi}_i &= - \frac{\partial E}{\partial \phi_i} (\phi; \theta) \\
        &= - \sum_j W_{ij} \sin(\phi_i - \phi_j) - h_i \sin(\phi_i - \psi_i).
    \end{split}
\end{equation}
It is obvious from this equation that we took $E$ (as well as the couplings and the biases) to have the dimensions of a frequency, to avoid having to introduce a separate mobility constant in front of the gradient.  These are the equations of motion for a standard  Kuramoto model \Citep{KuramotoReview_Acebron} modified by the bias term $\sum_i h_i \cos(\phi_i-\psi_i)$. In this sense, Eq.~\eqref{eq:EoM} describes the dynamics of a perfectly synchronized network of oscillators with the same intrinsic frequency. Here and in the following, $\theta$ denotes the set of all the trainable parameters---in our case, $\theta = (W, h, \psi)$. As explained in the introduction, such a system can be realized in many physical platforms.

 \subsection{Equilibrium Propagation}
 
        EP was introduced in~\Citep{scellier2017equilibrium,scellier2021deep} as a way to perform physical backpropagation, i.e., to extract training gradients based on the physical dynamics of a system undergoing relaxation towards a minimum-energy state. In a nutshell, it involves relaxing the system under two different choices of boundary conditions (``free'' and ``nudge''), comparing the resulting states locally and performing parameter updates via some feedback mechanism (see below). As such, EP stands in the tradition of contrastive learning approaches which have been a well-known part of machine learning since the 80s \Citep{ackley1985learning}, but importantly EP offers an implementation of those general mathematical ideas in terms of local physical interactions. In recent years, EP has been explored further in various directions.  A variant termed ``coupled learning'' has been introduced~\Citep{stern2021supervised} (see \Citep{scellier2023energybased} for a very useful comparison of the different versions of such energy-based training approaches). EP has been proposed for the training of nonlinear resistor networks~\Citep{kendall2020training}, and it has also been extended to deal with spiking networks~\Citep{martin2021eqspike}. The effects of the EP/ coupled-learning training  dynamics on the response behaviour of a neuromorphic system was analyzed carefully in~\Citep{stern2023physical}. In addition, for EP, it was realized that it is possible to implement continual parameter updates already during the second phase of the equilibration dynamics, leading to an approach that is local in time~\Citep{ernoult2020equilibrium}. Going an important step further, recently, a dynamical version of EP has been proposed~\Citep{scellier2022agnostic} (see also~\Citep{falk2023contrastive}), where in addition to gradient extraction via physical dynamics there is also a way to perform the parameter update itself via physical dynamics, rather than via measurement and feedback. This development can be seen in the conceptual tradition of Hamiltonian Echo Backpropagation~\Citep{lopez2021self}, where a general way to perform training updates via physical dynamics in neuromorphic systems was first introduced.

        On the experimental side, EP and its variants have been implemented so far in only a few devices---almost all of them electronic systems. One set of demonstrations investigated linear and nonlinear resistor networks in a platform using two copies of the network working in tandem, trained using coupled learning~\Citep{dillavou2022demonstration,wycoff2022desynchronous,stern2022physical,dillavou2023machine}. An experimental application to elastic networks was recently presented in~\Citep{altman2023experimental}. Furthermore, a classical Ising model has been trained using the ideas of EP and making use of a quantum annealer to efficiently reach equilibrium~\Citep{laydevant2023training}. Probably the most technically advanced platform consists of a memristor crossbar array trained using a contrastive learning approach closely related to EP~\Citep{yi2023activity,oh2023memristor}.
        
        In general, EP is applied to energy-based models, i.e., systems with an energy function. We will now describe its workings, already keeping in mind the system of coupled phase oscillators. During evaluation (inference), the input is injected by fixing the phases of some selected oscillators to the components of the input vector, e.g., representing the color values of a picture. This could as well be done by (sufficiently strong) external driving, which will generate the same tendency towards phase locking already present in the bias terms. The system then equilibrates via its natural dynamics, i.e., it relaxes to the lowest-energy state. The output is finally read off by observing a selected set of oscillators and taking their phases to represent the output vector. 
        
        We call the vector of input phases $\phi_{\rm in}$, while the phase angles of the output units and the hidden units are denoted as $\phi_{\rm out}$ and $\phi_{\rm hidden}$, respectively. The energy function can then be written as $E(\phi_{\rm in},\phi_{\rm hidden},\phi_{\rm out};\theta)$. In the context of EP, we call this 
        the \textbf{internal energy}.

       During training, we want to measure the deviation of the actual output to the desired output (the target). In the context of phase oscillators, a simple possibility is the following \textbf{distance function}. It simply compares the normalized spin vectors $\bold{s}$, leading to a cosine-similarity measure, with the desired output values (target) denoted as $\phi_i^\tau$:
        \begin{equation}\label{eq:similarityFunction}
    	    D(\phi, \phi^\tau) \equiv \sum_{i \in O} \frac{1}{2} | \bold{s}_i - \bold{s}_i^\tau |^2 = \sum_{i \in O} 1- \cos(\phi_i - \phi_i^\tau ).
    	\end{equation}
       Training via EP requires the system to have another weak contribution to the energy function which can be switched on when needed. This component describes the interaction between the output and the desired target. It plays the role of a cost function in supervised learning and is called \textbf{external energy}.  The most straightforward ansatz for the cost function would be to employ the distance function introduced above. However, we found that this generates problems, since even incorrect solutions (for which phases differ by $\pi$) are fixed points---albeit unstable. Empirically, the following cost function works much better, as it eliminates these undesired unstable fixed points and quickly drives the system away from incorrect solutions
        \begin{equation}\label{eq:cost_function}
            C(\phi, \phi^\tau) = \sum_{i \in S_{\rm out}} -\ln(1+ \cos(\phi_i - \phi_i^\tau) ).
        \end{equation}

We  define  $S_{\rm in}$, $S_{\rm h}$ and $S_{\rm out}$ as the sets of input, hidden and output units, respectively. Although the logarithm appearing here was introduced empirically, it can be motivated by the connection to the categorical cross-entropy (Kullback-Leibler divergence) found for classification tasks. Interestingly, the cost function adopted here can also be related to fidelities occuring when comparing different quantum states, despite the classical setting we are working in, as shown in Appendix \Ref{Appendix:interpretation}. 

Once a cost function has been assigned, EP introduces a small parameter $\beta$ which quantifies the strength of the interaction between the output units and the target. This leads to the \textbf{total energy} of the system
\begin{equation}
    F(\phi,\theta,\beta) = E(\phi_{\rm in},\phi_{\rm hidden},\phi_{\rm out};\theta) + \beta C(\phi_{\rm out}, \phi^\tau).
\end{equation}
        
The process of EP \cite{scellier2017equilibrium} contains the following steps: 
\begin{enumerate}
    \item Initialize the system such that the input units are fixed to the provided input values.
    \item Set $\beta=0$ and obtain a stable fixed point/ equilibrium by following the relaxation dynamics 
    \begin{equation}
        \dot{\phi_i} = -\frac{\partial F}{\partial \phi_i}(\phi,\theta,\beta=0)
    \end{equation}
    towards the fixed point. This state is called the \textbf{free equilibrium} and this step is called the \textbf{free phase} of EP. 
    \item  Switch on a small value of $\beta$ and again follow the dynamics towards a new fixed point. This state is called \textbf{nudge equilibrium} and this step is called the \textbf{nudge phase} of EP.
    \item Repeat the process 1-3 for all the input data from the given batch of training samples. The central insight of EP \Citep{scellier2017equilibrium} is that the gradient of the cost function with respect to the trainable parameters can be approximated as
    \begin{equation}\label{eq:EPgradient}
        \begin{split}
            \frac{\partial C}{\partial \theta_\alpha}  &= \frac{d}{d \beta} \left( \frac{\partial F}{\partial \theta_\alpha} \right)_{\beta = 0} \\
         &\approx \frac{1}{\beta}\left( \left\langle {\partial E \over \partial \theta_\alpha} \right\rangle^{\rm nudge}_{\phi_{\rm in}} - \left\langle {\partial E \over \partial \theta_\alpha} \right\rangle^{\rm free}_{\phi_{\rm in}} \right).
        \end{split}
    \end{equation}
    in which, in our case, ${\partial E \over \partial \theta_\alpha}$ are given by the following analytical expressions
    \begin{equation}
         \begin{split}
             {\partial E \over \partial W_{ij}} &= -\cos (\phi_i - \phi_j) \\
             {\partial E \over \partial h_i} &= -\cos (\phi_i - \psi_i) \\
             {\partial E \over \partial \psi_i} &= - h_i \sin(\phi_i - \psi_i).
         \end{split}
    \end{equation}
        \end{enumerate}    
    It has been discovered that EP gradients may benefit from nudging the system away from the desired target state, i.e. $\beta<0$, or even estimating the gradient in a symmetric way from positive and negative nudging \Citep{laborieux2021scaling}. However, it was not possible to use this trick in our work due to the fact that our cost function can become large quickly when nudged in the 'wrong' direction.
    We implement the dynamics of the equilibration numberically using the 4th order Runge-Kutta method with adaptive step size, implemented in python/ jax on a GPU.

        \begin{figure}[htbp]
    	    \centering
    	    \includegraphics[width = 0.5 \textwidth]{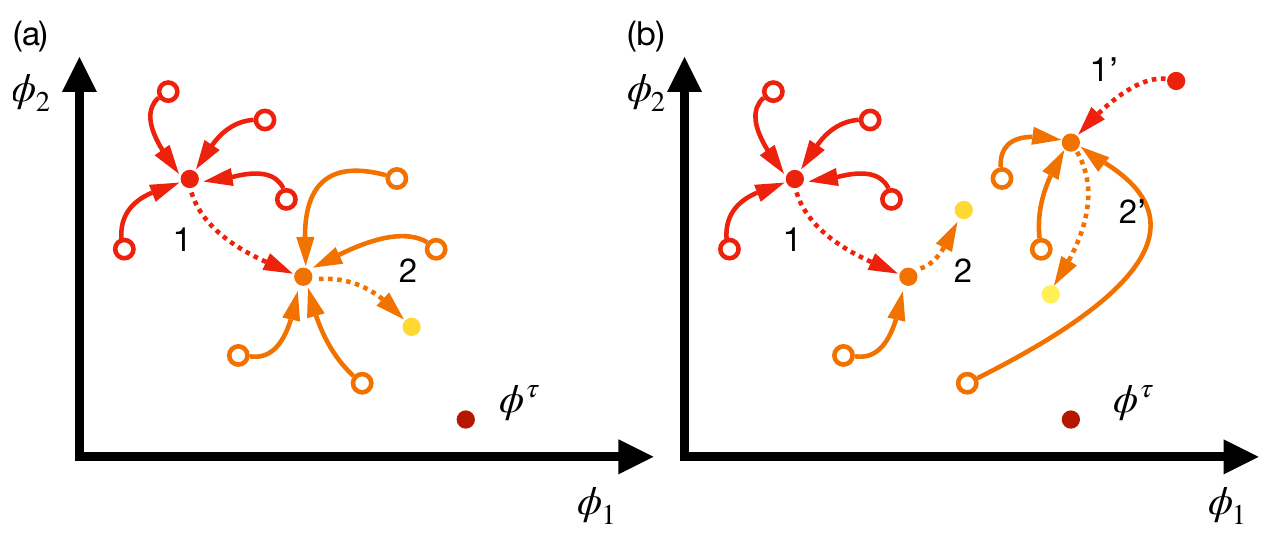}
    	    \caption{\textbf{The problem of multi-stability.}
    	    (a) A case with only one stable fixed point. The circles refer to the initial states and the arrows refer to the relaxation to the stable fixed points (equilibria; solid dots).  The arrows with dashed lines refer the evolution of the equilibria during the training. The dark red dot refers to the target output, assuming that the two phase coordinates shown here both refer to output units, i.e., hidden phases and input phases are suppressed in this depiction. Different colors refer to different iterations during the training. 
            (b) Multistability. Different equilibria can be reached at the same training iteration (same color), starting from different random initial conditions.
    	    }
    	    \label{fig:multi-stability}
    	\end{figure}%

        As we focus on the deterministic dynamics of a system with a complicated non-convex energy landscape, a discussion of multistability is inevitable. Apparently such a discussion was not needed in earlier applications of EP, due to the simpler nature of the energy landscape encountered there (e.g.~\Citep{kendall2020training}). An XY model with random couplings can be viewed as a spin glass, for which \Citep{S_F_Edwards_1976} shows that the number of fixed points grows exponentially with the  size of the network. Although only a small fraction of these fixed points are stable, the effect of possible multistability in this energy landscape cannot be neglected, as it can undermine the training if it is not properly handled. We emphasize that on top of this multistability of the energy landscape (peculiar to energy-based models) which hinders equilibration for each training step, there can also be multistability in the cost function training landscape (similar to neural networks), which is a distinct aspect.

        The effect of multi-stability is shown schematically in Fig.~\Ref{fig:multi-stability}.  In panel~(a), we assume that the system has only one stable fixed point. The fixed point then moves along the dashed arrows and approaches the desired result (the target) by adjusting the trainable parameters with EP. However, if there are multiple stable fixed points, the situation becomes more complicated, as shown in panel (b). Depending on the arbitrarily chosen initial phase configuration one may end up in different fixed points. Depending on which training updates are performed in which fixed points, this may even result in situations where some of the fixed points move away from the target during training, at least temporarily. Sometimes there are even more drastic cases as a stable fixed point may become unstable after the parameters have been adjusted. These cases can reduce the efficiency of the training. 
        
       In principle, the multistability problem can be solved by introducing noise into the system and rewriting Eq.~\eqref{eq:EPgradient} with a Boltzmann distribution (the theory is briefly discussed in Ref.~\Citep{scellier2021deep}). However, we note here that the challenge of multistability can also be solved within the framework of deterministic dynamics. We propose to randomly initialize the hidden (and output) units for each new training step. The system will correspondingly end up in different fixed points, depending on the location of the randomly selected initial conditions with respect to the basins of attraction of those fixed points.  Then, in the spirit of Stochastic Gradient Descent (SGD), we average the learning gradient over both the random initial configurations and the training samples. Based on the success of SGD in other contexts, we expect that all stable fixed points can be trained simultaneously (and our numerical experiments confirm this). 
       
       In practice, in each iteration we randomly initialize the system many times and average the approximate gradient given by EP over both the training sample batch and the initial states. We denote the number of random initial values by $M_{\rm init}$ and the size of the training batch by $M_{\rm data}$, and we update the trainable parameters accordingly
        \begin{equation}\label{eq:update_rule}
    	    \begin{split}
    	        \delta \theta_\alpha = & -\frac{\eta}{\beta} \frac{1}{M_{\rm data}} \sum_{i=1}^{M_{\rm data}} \frac{1}{M_{\rm init}} \sum_{j=1}^{M_{\rm init}}  \\ 
                &\left( \frac{\partial E}{\partial \theta_\alpha}(\phi^{i,j ,\rm nudge}; \theta)  - \frac{\partial E}{\partial \theta_\alpha}(\phi^{i,j ,\rm free}; \theta) \right).
    	    \end{split}
    	\end{equation}
        Here $\phi^{i,j ,\rm nudge/free}$ refers to the nudge/ free equilibrium for the $i$th input attained from the $j$th initial state, and $\eta$ refers to the learning rate.

	\section{Results}
	    \subsection{Learning the XOR function}
        We first test EP-based training of a coupled-oscillator network for the simple case of the XOR function. For artificial neural networks, the XOR problem is historically important, because it was the first task to require at least one hidden layer. Due to its simplicity, the XOR task allows us to investigate the dynamical properties of the network in great detail. We are able to explore the effects of the network size, the evolution of equilibria when the network is multi-stable, and the influence of random initial configurations. 

        \begin{figure}[htbp]
    	    \centering
    	    \includegraphics[width = 0.5 \textwidth]{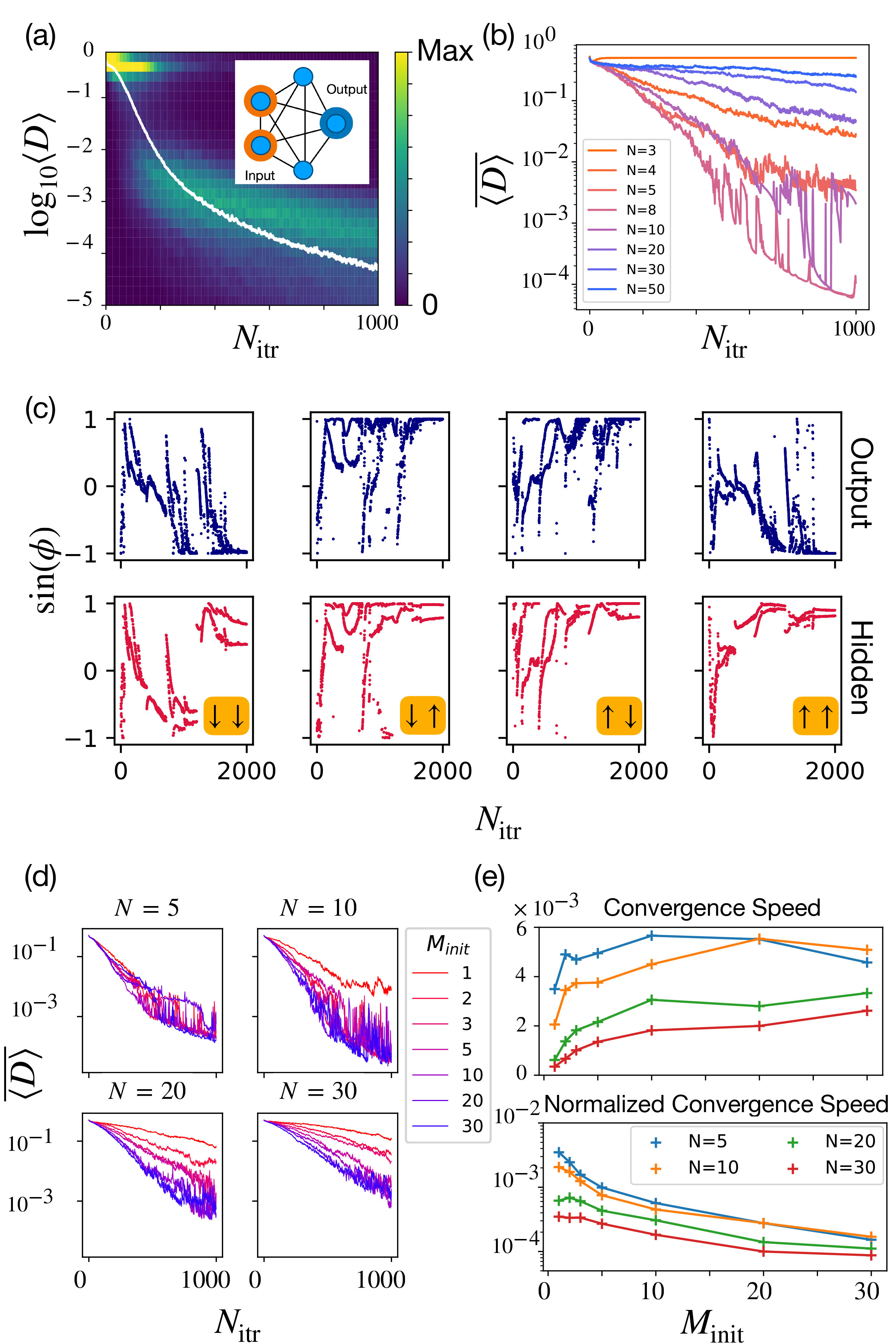}
    	    \caption{\textbf{Results of training an XY model on the XOR task.} (a)~Statistics of the training progress of 5-unit XY networks of all-to-all connectivity (the structure is shown in the upper right corner). Evolution of mean distance function $\langle D \rangle$ (white line, average over 1000 training runs). The density plot indicates the histogram of $\log_{10}D$ over these runs. (b)The dependence of the learning speed on the size $N$. For a given $N$, we randomly initialize the trainable parameters and train them for 1000 iterations, repeating this whole process 100 times. The curves show the evolution of  $\overline{\langle D \rangle}$ during the training. (c)~Evolution of equilibria during a single training run. For each of the four input configurations, the possible stable fixed points at each training iteration are found by running the system with random initialization for 100 times. For each of these fixed points the state of the output unit (upper row) and one of the hidden units (lower low) are recorded. Output units for all fixed points converge to a unique, deterministic function of the input at the end of training. (d)~Dependency of learning speed on the number of random initial configurations, $M_{\rm init}$
            For each $N$ and $M_{\rm init}$ we show the evolution of $\overline{\langle D \rangle}$ averaged over 100 training runs. (e)~Upper panel: the learning speed of training measured by the average slope of the first 300 steps shown in (d) and its dependence on $M_{\rm init}$. Lower panel: the ``physical training speed'', measured by the slope divided by $M_{\rm init}$. }
    	    \label{fig:XOR}
	    \end{figure}%

            We encode ``True'' and ``False'' with $\phi=\pi/2$ and $\phi=-\pi/2$, respectively. For XOR,  if the two inputs are same, the output unit should give us a ``False'' and vice versa. 

            At the beginning of training, the weights are initialized independently according to a standard Gaussian distribution, while the strength and the angle of the bias are initialized according to the uniform distribution over $[-0.5, 0.5)$ and $[-\pi, \pi)$, respectively. During the training, we repeatedly update the parameters according to Eq.~\eqref{eq:update_rule}. In each iteration, the system is exposed to all 4 input-output pairs ($M_{\rm data} = 4$). For each such pair, we repeat the following process for $M_{\rm init}$ times: the units of the network are initialized so that the phase angles of the hidden and output units independently follow the uniform distribution over $[-\pi, \pi)$; then we solve Eq.~\eqref{eq:EoM} for a fixed time interval $T$ which is sufficiently long to ensure that the coupled phase oscillators approximately reach their equilibrium; finally, we calculate the average parameter gradient and update the trainable parameters using Eq.~\eqref{eq:update_rule}.  In this section, we fix $T=100$ and $\eta=0.1$. 
            
            All our numerical results for the XOR task are collected in Fig.~\ref{fig:XOR}.  We first set the number of initial configurations to $M_{\rm init}=1$ and train a 5-unit XY network with all-to-all connectivity for 1000 iterations. In each iteration, the mean distance  $\langle D \rangle$ averaged over the 4 input-output pairs (computed with Eq.~\eqref{eq:similarityFunction}) is recorded. In addition, we perform averages over different training runs with random initial weight configurations. The average distance is shown in panel (a) of Fig.\ref{fig:XOR}. The white line depicts the average mean distance, which we define to be $\overline{\langle D \rangle} = \frac{1}{N_{\rm train}}\sum_{i=1}^{N_{\rm train}} \langle D(\phi, \phi^{\tau}) \rangle$, where the average is taken over training runs. We find that the distance almost always converges to 0, which proves the ability of XY networks to learn XOR via EP training. 

            The effect of size $N$ on learning speed is shown in panel (b). For different $N$, we initialize and train an all-to-all network 100 times and plot the evolution of the mean distance function $\overline{\langle D \rangle}$. The results in (b) show that an increasing size initially accelerates the learning process. The speed of learning reaches its maximum at a certain number of phase oscillators and then decreases. The initial increase in speed can be explained by the increasing number of parameters, which makes the network more flexible. However, when the size becomes excessively large, the large number of parameters and multiple stable fixed points (all of which have to be trained) will reduce the efficiency of the training. According to the result in (b), the training becomes slower again when the number of units exceeds 8. 

           We now turn to a more detailed investigation of multistability [panel (c) of Fig.\Ref{fig:XOR}]. We train a network of 30 units with 2000 iterations and record the parameters for each iteration. After training, for each iteration and each input-output pair, we search for all the possible equilibria by randomly initializing and running the network 100 times. We then inspect the phase angles for each of these equilibria. In the top row of panel (c), we show the evolution of the set of phase angles for an output unit. Importantly, training converges in the sense that the output unit phase angle becomes a deterministic function of the input, with no remaining scatter. At the same time however, multistability in the energy landscape remains. We can see this by observing the evolution of one selected hidden unit (bottom row), which shows scatter even after conclusion of the training. Overall, we confirm that despite the persistence of multistability, it is still possible to adjust all of the equilibria simultaneously to perfectly attain the training objective.

            Finally, we study the effect of the number of random initial configurations, $M_{\rm init}$, considered during each training iteration. First, for a fixed number of units $N$ and $M_{\rm init}$, we train an XY network with 100 different sets of initial trainable parameters and record the average mean distance function $\overline{\langle D \rangle}$. The results are shown in panels (d) and (e). From panel (d), we observe that the learning speed increases with increasing $M_{\rm init}$ and saturates at large $M_{\rm init}$. The effect of $M_{\rm init}$ is not obvious for networks of 5 units, as multi-stability is rare for small networks. 

            Another observation is that the average mean distance function $\overline{\langle D \rangle}$ always decays approximately exponentially in the early stages of training. Therefore, we perform a linear regression to $\log \overline{\langle D \rangle}$ on the first 300 iterations and use the negative slope to measure the speed of learning. The results are shown in the upper graph of panel (e). An increase in the learning speed and a saturation at large $M_{\rm init}$ are evident.  However, a larger $M_{\rm init}$ physically means more equilibration runs. This results in larger time consumption in the physical world. To take this into account, we divide the learning speed extracted above by $M_{\rm init}$ and use this ratio to measure the physical speed of learning (``normalized convergence speed'' in the figure). The results are shown in the bottom graph. We observe that the physical learning speed reaches its maximum at a relatively small value of $M_{\rm init}$. This indicates that a small number of different initial configurations can already sufficiently accelerate the learning process in terms of reducing the physical time consumption.

	\subsection{Handwritten-Digits Recognition}

        \begin{figure}[h!]
        	\centering
        	\includegraphics[width = 0.5 \textwidth]{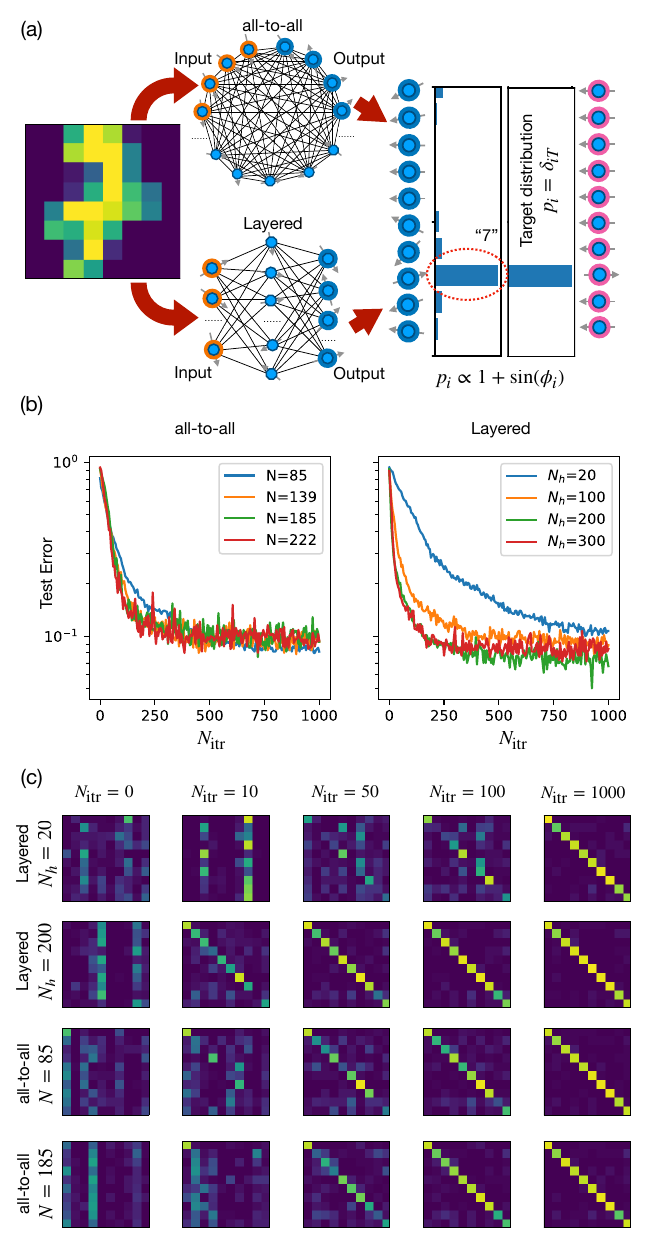}
        	\caption{\textbf{Handwritten-digit recognition in a network of coupled phase oscillators. }(a): We consider both all-to-all and layered connectivity networks, where the recognized digit is one-hot encoded in the phase configuration of the 10 output units. (b): Training curves for networks of different sizes with all-to-all connectivity and layer structure. The sizes of the networks are selected such that the number of trainable parameters for them are approximately the same (shown in \Ref{tab:network_paras}) (d): Training evolution of confusion matrices, for different structures and different network sizes.
            }
        	\label{fig:mnist}
    	\end{figure}%

        In the previous section we showed the ability of XY networks to learn the simple XOR task via EP. In this section, we will analyse the performance on the recognition of written digits, a more complex task which has long been considered a benchmark for neural networks. 

            To keep the numerical effort in our simulations manageable, we use a scaled-down version of the MNIST dataset. Specifically, we use the dataset provided by ScikitLearn which is a copy of the test set for the Optical Recognition of Handwritten Digits from the UCI ML hand-written digits datasets \Citep{misc_optical_recognition_of_handwritten_digits_80}. It consists of 1797 8$\times$8-pixel greyscale images of written digits, approximately 180 for each. A sample figure is shown in the left of panel (a) of Fig \ref{fig:mnist}. The size of the data set and the dimension of the pictures are all smaller than the commonly used MNIST, which allows us to do training via EP in smaller coupled-oscillator networks and test various architectures more easily. The structure of the networks and the results are shown in Fig \ref{fig:mnist}.

            As shown in Fig \ref{fig:mnist}(a), we train XY networks with both all-to-all connectivity and a layered structure. In each such network, there are 64 input units and 10 output units. The input data (image pixels) are linearly rescaled into the range $[-\pi/2, \pi/2]$. As usual for classification tasks, the outputs are interpreted as a probabilities, such that $p_i \propto 1+\sin \phi_i$, as shown in the right side of panel (a), where $p_i$ denotes the probability that ``the input digit is $i$''. The inference result is given by taking the index of the highest probability. We note that in our setting these probabilities are not normalized, i.e. they do not sum up to one. This is different from usual artificial neural networks, where one uses a softmax activation function in the last layer to ensure normalization. The label for each input image is encoded in a ``one-hot'' configuration, giving a phase of $\pi/2$ at the output unit (phase oscillator) corresponding to the correct label of the digit seen in the image and $-\pi/2$ otherwise.
            We use the same cost function as for the XOR task. 
            
            For neural networks of large size, the influence of weight initialization on training can be crucial\Citep{glorot2010understanding}. In this section, we initialize out networks as following: 
            \begin{enumerate}
                \item For all-to-all networks, the weights are initialized according to $\mathcal{N}(0, 1/\sqrt{N})$, where $\mathcal{N}$ refers to a normal Gaussian distribution and $N$ is the total number of units in the network. The strength of the initial bias field is set to be 0 everywhere, while the initial bias directions are randomly set according to a uniform distribution over $[-\pi, \pi]$. 
                \item For layered networks, the initial weights connecting the $i$th and the $i+1$th layers are initialized according to $\mathcal{N}(0, 1/\sqrt{N_{i}+N_{i+1}})$, where $N_i$ and $N_{i+1}$ are the size of the $i$th and the $i+1$th layer. The biases are initialized in the same way as for all-to-all networks. 
            \end{enumerate}
            In order to evaluate the effect of the architecture, we compare the training process of all-to-all and layered XY networks with a similar number of parameters. For an all-to-all network of $N$ units containing $N_{in}$ input units and $N_{out}$ output units, the number of trainable parameters (weights, strength and direction of bias) is $\frac{1}{2} N(N-1) - \frac{1}{2} N_{in}(N_{in}-1) + 2(N - N_{in})$. For a network of layered structure, the number of trainable parameters is $\sum_{i=0}^{L-1} N_i N_{i+1} + 2\sum_{i=1}^{L} N_i$, where $L$ here denotes the depth of the network (number of layers). The focus of our work is on the performance of all-to-all networks and layered networks with one hidden layer. The sizes and parameters of the tested networks are displayed in Tab.~\ref{tab:network_paras}.
            
            \begin{table}[htbp]
                \centering
                \begin{tabular}{ |p{2cm}|p{2cm}|p{2cm}|p{2cm}|  }
                     \hline
                     \multicolumn{2}{|c|}{Layered} &
                     \multicolumn{2}{|c|}{All-to-All} \\
                     \hline
                     Layer Structure & \# Parameters & \# Units & \# Parameters\\
                     \hline
                     
                     64, 20, 10   & 1540    & 85 &   1596 \\
                     64, 100, 10 &   7620  & 139   & 7725 \\
                     64, 200, 10 & 15220 & 185 &  15246 \\
                     64, 300, 10    & 22820 & 222 &  22831 \\
                     
                     \hline
                \end{tabular}
                \caption{Sizes and parameters of tested networks}
                \label{tab:network_paras}
            \end{table}%

                The results are shown in panels (b) and (c) of Fig.~\ref{fig:mnist}. The data set is split into a training set and a test set. The training set consists of the first 100 images for each of the 10 digits (1000 images in total), while the test set consists of the next 70 images for each digit. All networks are trained for 1000 iterations. In each iteration, we randomly select 30 images from the training set for each digit (so the total batch size is 300). We take the number of random initial phase configurations for each sample, $M_{\rm init}$, to be one. This can be justified by our insights from the XOR task [panel (f) in Fig.~\ref{fig:XOR}] and the fact that a large batch size, such as the one adopted here, automatically leads to a variety of different initial configurations. Every 5 iterations, we test the accuracy of the inference with all the images in the test set. 

                The evolution of the test errors is shown in Fig.~\ref{fig:mnist}~(b) and we list the best attained test accuracy for each architecture in Tab.~\ref{tab:accuracies} of Appendix~\ref{sec:accuracies} in comparison to results we obtained with artificial neural networks (ANNs) and linear classifiers.
                Increasing the number of parameters (by increasing the number of hidden units) seems to improve the test accuracy for the layer architecture although the accuracy again seems to drop as we increase the number of hidden units from $200$ to $300$.
                The best test accuracy we obtained with the layer architecture (at $200$ hidden units) was $94.1\,\%$ which is above the accuracy of $90.7\,\%$ achieved by a linear classifier with the same number of parameters but below the accuracy of $95.0\,\%$ achieved by an ANN with one hidden layer (see Appendix~\ref{sec:accuracies} for further details). In contrast, the performance of a network with all-to-all connectivity remains approximately constant as we increase the number of parameters with a slight drop as the number of hidden units increases from $11$ to $75$.
                The best accuracy we achieved with all-to-all connectivity was $93.3\,\%$ with $11$ hidden units. The comparable linear classifier achieves $90.4\,\%$ accuracy, while the ANN attains $94.3\,\%$.
                This indicates that EP can be used successfully to train a network of coupled phase oscillators to perform standard classification tasks. The performance does not reach the state-of-the-art values of artificial neural networks, which have been highly optimized over the past two decades for the MNIST task (reaching accuracies close to 100\% on the larger $28\times 28$  images) , but this was to be expected – further optimization and evolution of the neuromorphic platform (as well as the details of the training procedure) would certainly yield improved accuracy. In addition, we recall that the motivation for adopting neuromorphic platforms is to achieve gains in energy efficiency and other aspects like inference speed. Therefore, one may be willing to pay the price of reduced accuracy in certain applications.
                
                The influence of the network architecture on convergence is also evident. For all-to-all networks, the curves depicting the training evolution of the test error almost overlap for the different choices of network size. In contrast, for layered networks, convergence speed increases significantly as the size of the hidden layer increases. However, as the size of the hidden layer increases above 200, the speed of convergence tends to saturate. This difference suggests that the speed of learning is more sensitive to the size of the networks when they are layered.
    
                Finally, we visualize the training evolution via the confusion matrix, which measures the probability that a given true digit is classified as one of the ten possible digits (correct classifications are on the diagonal). For different network sizes, we plot the confusion matrix after 0, 10, 50, 100 and 1000 iterations during training. Comparing the results of the same network architecture, we see a clear influence of size on the learning speed for layered networks, which is negligible for all-to-all networks. Comparing the the results for the a similar number of trainable parameters, we find that the learning process of layered networks is slower at smaller sizes, but faster when the size is sufficiently large. 

                In terms of network architectures, it is a natural question to ask whether a locally coupled network can also yield good training performance. This is an important question, since physical interactions are naturally local, so a neuromorphic device based on such an architecture would require less physical resources. We have explored this question and investigated nearest-neighbor-coupled square lattices of phase oscillators. In that setting, there are various choices to be made, e.g. as to how the input and output units are distributed over the lattice. In any case, despite our best efforts, we could not make that architecture train successfully on the MNIST task. Intuitively, we believe this is due to the difficulty for the system to find ways to effectively process different parts of the input that are spatially removed from each other (and from the output units). This difficulty may also be related to the physics of localization in disordered systems. More detailed investigations here would be needed in the future, since the performance of locally coupled neuromorphic devices in non-trivial tasks is an important question, even beyond the coupled phase oscillator platform discussed here.
	
	\section{Conclusions}

    Our results confirm that it is possible to employ equilibrium propagation for training systems of coupled phase oscillators. As an important aspect, we observed that the complex energy landscape of the XY model leads to multistability. We showed that it is possible to address that challenge via stochastic initialization, leading to a simultaneous adjustment of all fixed points and eventually stable convergence of the training. In terms of physical implementations, we feel the most promising route might be based on coupled laser arrays of the type shown in \Citep{nixon2013observing}, possibly combined with the flexibility of spatial light modulators or similar devices for setting the coupling weights, as shown for photonic neural networks in \Citep{bueno2018reinforcement}. However, many other platforms are conceivable as well, with the main requirements that it is possible to read out phases, drive individual oscillators externally (for input injection and biases), as well as implement tuneable couplings between arbitrary oscillators.

\section{Acknowledgements}
    We acknowlegde funding via the German Research Foundation
    (DFG, Project-ID 429529648–TRR 306 QuCoLiMA, “Quantum Cooperativity of Light and Matter”).

	\bibliography{refs} 

\begin{thebibliography}{70}%
\makeatletter
\providecommand \@ifxundefined [1]{%
 \@ifx{#1\undefined}
}%
\providecommand \@ifnum [1]{%
 \ifnum #1\expandafter \@firstoftwo
 \else \expandafter \@secondoftwo
 \fi
}%
\providecommand \@ifx [1]{%
 \ifx #1\expandafter \@firstoftwo
 \else \expandafter \@secondoftwo
 \fi
}%
\providecommand \natexlab [1]{#1}%
\providecommand \enquote  [1]{``#1''}%
\providecommand \bibnamefont  [1]{#1}%
\providecommand \bibfnamefont [1]{#1}%
\providecommand \citenamefont [1]{#1}%
\providecommand \href@noop [0]{\@secondoftwo}%
\providecommand \href [0]{\begingroup \@sanitize@url \@href}%
\providecommand \@href[1]{\@@startlink{#1}\@@href}%
\providecommand \@@href[1]{\endgroup#1\@@endlink}%
\providecommand \@sanitize@url [0]{\catcode `\\12\catcode `\$12\catcode
  `\&12\catcode `\#12\catcode `\^12\catcode `\_12\catcode `\%12\relax}%
\providecommand \@@startlink[1]{}%
\providecommand \@@endlink[0]{}%
\providecommand \url  [0]{\begingroup\@sanitize@url \@url }%
\providecommand \@url [1]{\endgroup\@href {#1}{\urlprefix }}%
\providecommand \urlprefix  [0]{URL }%
\providecommand \Eprint [0]{\href }%
\providecommand \doibase [0]{http://dx.doi.org/}%
\providecommand \selectlanguage [0]{\@gobble}%
\providecommand \bibinfo  [0]{\@secondoftwo}%
\providecommand \bibfield  [0]{\@secondoftwo}%
\providecommand \translation [1]{[#1]}%
\providecommand \BibitemOpen [0]{}%
\providecommand \bibitemStop [0]{}%
\providecommand \bibitemNoStop [0]{.\EOS\space}%
\providecommand \EOS [0]{\spacefactor3000\relax}%
\providecommand \BibitemShut  [1]{\csname bibitem#1\endcsname}%
\let\auto@bib@innerbib\@empty
\bibitem [{\citenamefont {Markovi{\'c}}\ \emph {et~al.}(2020)\citenamefont
  {Markovi{\'c}}, \citenamefont {Mizrahi}, \citenamefont {Querlioz},\ and\
  \citenamefont {Grollier}}]{markovic2020physics}%
  \BibitemOpen
  \bibfield  {author} {\bibinfo {author} {\bibfnamefont {D.}~\bibnamefont
  {Markovi{\'c}}}, \bibinfo {author} {\bibfnamefont {A.}~\bibnamefont
  {Mizrahi}}, \bibinfo {author} {\bibfnamefont {D.}~\bibnamefont {Querlioz}}, \
  and\ \bibinfo {author} {\bibfnamefont {J.}~\bibnamefont {Grollier}},\
  }\href@noop {} {\bibfield  {journal} {\bibinfo  {journal} {Nature Reviews
  Physics}\ }\textbf {\bibinfo {volume} {2}},\ \bibinfo {pages} {499} (\bibinfo
  {year} {2020})}\BibitemShut {NoStop}%
\bibitem [{\citenamefont {Christensen}\ \emph {et~al.}(2022)\citenamefont
  {Christensen}, \citenamefont {Dittmann}, \citenamefont {Linares-Barranco},
  \citenamefont {Sebastian}, \citenamefont {Le~Gallo}, \citenamefont
  {Redaelli}, \citenamefont {Slesazeck}, \citenamefont {Mikolajick},
  \citenamefont {Spiga}, \citenamefont {Menzel} \emph
  {et~al.}}]{christensen20222022}%
  \BibitemOpen
  \bibfield  {author} {\bibinfo {author} {\bibfnamefont {D.~V.}\ \bibnamefont
  {Christensen}}, \bibinfo {author} {\bibfnamefont {R.}~\bibnamefont
  {Dittmann}}, \bibinfo {author} {\bibfnamefont {B.}~\bibnamefont
  {Linares-Barranco}}, \bibinfo {author} {\bibfnamefont {A.}~\bibnamefont
  {Sebastian}}, \bibinfo {author} {\bibfnamefont {M.}~\bibnamefont {Le~Gallo}},
  \bibinfo {author} {\bibfnamefont {A.}~\bibnamefont {Redaelli}}, \bibinfo
  {author} {\bibfnamefont {S.}~\bibnamefont {Slesazeck}}, \bibinfo {author}
  {\bibfnamefont {T.}~\bibnamefont {Mikolajick}}, \bibinfo {author}
  {\bibfnamefont {S.}~\bibnamefont {Spiga}}, \bibinfo {author} {\bibfnamefont
  {S.}~\bibnamefont {Menzel}},  \emph {et~al.},\ }\href@noop {} {\bibfield
  {journal} {\bibinfo  {journal} {Neuromorphic Computing and Engineering}\
  }\textbf {\bibinfo {volume} {2}},\ \bibinfo {pages} {022501} (\bibinfo {year}
  {2022})}\BibitemShut {NoStop}%
\bibitem [{\citenamefont {Prezioso}\ \emph {et~al.}(2015)\citenamefont
  {Prezioso}, \citenamefont {Merrikh-Bayat}, \citenamefont {Hoskins},
  \citenamefont {Adam}, \citenamefont {Likharev},\ and\ \citenamefont
  {Strukov}}]{prezioso2015training}%
  \BibitemOpen
  \bibfield  {author} {\bibinfo {author} {\bibfnamefont {M.}~\bibnamefont
  {Prezioso}}, \bibinfo {author} {\bibfnamefont {F.}~\bibnamefont
  {Merrikh-Bayat}}, \bibinfo {author} {\bibfnamefont {B.~D.}\ \bibnamefont
  {Hoskins}}, \bibinfo {author} {\bibfnamefont {G.~C.}\ \bibnamefont {Adam}},
  \bibinfo {author} {\bibfnamefont {K.~K.}\ \bibnamefont {Likharev}}, \ and\
  \bibinfo {author} {\bibfnamefont {D.~B.}\ \bibnamefont {Strukov}},\
  }\href@noop {} {\bibfield  {journal} {\bibinfo  {journal} {Nature}\ }\textbf
  {\bibinfo {volume} {521}},\ \bibinfo {pages} {61} (\bibinfo {year}
  {2015})}\BibitemShut {NoStop}%
\bibitem [{\citenamefont {Schneider}\ \emph {et~al.}(2018)\citenamefont
  {Schneider}, \citenamefont {Donnelly},\ and\ \citenamefont
  {Russek}}]{schneider2018tutorial}%
  \BibitemOpen
  \bibfield  {author} {\bibinfo {author} {\bibfnamefont {M.~L.}\ \bibnamefont
  {Schneider}}, \bibinfo {author} {\bibfnamefont {C.~A.}\ \bibnamefont
  {Donnelly}}, \ and\ \bibinfo {author} {\bibfnamefont {S.~E.}\ \bibnamefont
  {Russek}},\ }\href@noop {} {\bibfield  {journal} {\bibinfo  {journal}
  {Journal of Applied Physics}\ }\textbf {\bibinfo {volume} {124}} (\bibinfo
  {year} {2018})}\BibitemShut {NoStop}%
\bibitem [{\citenamefont {Shainline}\ \emph {et~al.}(2017)\citenamefont
  {Shainline}, \citenamefont {Buckley}, \citenamefont {Mirin},\ and\
  \citenamefont {Nam}}]{shainline2017superconducting}%
  \BibitemOpen
  \bibfield  {author} {\bibinfo {author} {\bibfnamefont {J.~M.}\ \bibnamefont
  {Shainline}}, \bibinfo {author} {\bibfnamefont {S.~M.}\ \bibnamefont
  {Buckley}}, \bibinfo {author} {\bibfnamefont {R.~P.}\ \bibnamefont {Mirin}},
  \ and\ \bibinfo {author} {\bibfnamefont {S.~W.}\ \bibnamefont {Nam}},\
  }\href@noop {} {\bibfield  {journal} {\bibinfo  {journal} {Physical Review
  Applied}\ }\textbf {\bibinfo {volume} {7}},\ \bibinfo {pages} {034013}
  (\bibinfo {year} {2017})}\BibitemShut {NoStop}%
\bibitem [{\citenamefont {Torrejon}\ \emph
  {et~al.}(2017{\natexlab{a}})\citenamefont {Torrejon}, \citenamefont {Riou},
  \citenamefont {Araujo}, \citenamefont {Tsunegi}, \citenamefont {Khalsa},
  \citenamefont {Querlioz}, \citenamefont {Bortolotti}, \citenamefont {Cros},
  \citenamefont {Fukushima}, \citenamefont {Kubota}, \citenamefont {Yuasa},
  \citenamefont {Stiles},\ and\ \citenamefont
  {Grollier}}]{TorrejonGrollier2017}%
  \BibitemOpen
  \bibfield  {author} {\bibinfo {author} {\bibfnamefont {J.}~\bibnamefont
  {Torrejon}}, \bibinfo {author} {\bibfnamefont {M.}~\bibnamefont {Riou}},
  \bibinfo {author} {\bibfnamefont {F.~A.}\ \bibnamefont {Araujo}}, \bibinfo
  {author} {\bibfnamefont {S.}~\bibnamefont {Tsunegi}}, \bibinfo {author}
  {\bibfnamefont {G.}~\bibnamefont {Khalsa}}, \bibinfo {author} {\bibfnamefont
  {D.}~\bibnamefont {Querlioz}}, \bibinfo {author} {\bibfnamefont
  {P.}~\bibnamefont {Bortolotti}}, \bibinfo {author} {\bibfnamefont
  {V.}~\bibnamefont {Cros}}, \bibinfo {author} {\bibfnamefont {A.}~\bibnamefont
  {Fukushima}}, \bibinfo {author} {\bibfnamefont {H.}~\bibnamefont {Kubota}},
  \bibinfo {author} {\bibfnamefont {S.}~\bibnamefont {Yuasa}}, \bibinfo
  {author} {\bibfnamefont {M.~D.}\ \bibnamefont {Stiles}}, \ and\ \bibinfo
  {author} {\bibfnamefont {J.}~\bibnamefont {Grollier}},\ }\href
  {http://arxiv.org/abs/1701.07715} {\bibfield  {journal} {\bibinfo  {journal}
  {CoRR}\ }\textbf {\bibinfo {volume} {abs/1701.07715}} (\bibinfo {year}
  {2017}{\natexlab{a}})},\ \Eprint {http://arxiv.org/abs/1701.07715}
  {1701.07715} \BibitemShut {NoStop}%
\bibitem [{\citenamefont {Wagner}\ and\ \citenamefont
  {Psaltis}(1987)}]{wagner1987multilayer}%
  \BibitemOpen
  \bibfield  {author} {\bibinfo {author} {\bibfnamefont {K.}~\bibnamefont
  {Wagner}}\ and\ \bibinfo {author} {\bibfnamefont {D.}~\bibnamefont
  {Psaltis}},\ }\href@noop {} {\bibfield  {journal} {\bibinfo  {journal}
  {Applied Optics}\ }\textbf {\bibinfo {volume} {26}},\ \bibinfo {pages} {5061}
  (\bibinfo {year} {1987})}\BibitemShut {NoStop}%
\bibitem [{\citenamefont {Shen}\ \emph {et~al.}(2017)\citenamefont {Shen},
  \citenamefont {Harris}, \citenamefont {Skirlo}, \citenamefont {Prabhu},
  \citenamefont {Baehr-Jones}, \citenamefont {Hochberg}, \citenamefont {Sun},
  \citenamefont {Zhao}, \citenamefont {Larochelle}, \citenamefont {Englund}
  \emph {et~al.}}]{shen2017deep}%
  \BibitemOpen
  \bibfield  {author} {\bibinfo {author} {\bibfnamefont {Y.}~\bibnamefont
  {Shen}}, \bibinfo {author} {\bibfnamefont {N.~C.}\ \bibnamefont {Harris}},
  \bibinfo {author} {\bibfnamefont {S.}~\bibnamefont {Skirlo}}, \bibinfo
  {author} {\bibfnamefont {M.}~\bibnamefont {Prabhu}}, \bibinfo {author}
  {\bibfnamefont {T.}~\bibnamefont {Baehr-Jones}}, \bibinfo {author}
  {\bibfnamefont {M.}~\bibnamefont {Hochberg}}, \bibinfo {author}
  {\bibfnamefont {X.}~\bibnamefont {Sun}}, \bibinfo {author} {\bibfnamefont
  {S.}~\bibnamefont {Zhao}}, \bibinfo {author} {\bibfnamefont {H.}~\bibnamefont
  {Larochelle}}, \bibinfo {author} {\bibfnamefont {D.}~\bibnamefont {Englund}},
   \emph {et~al.},\ }\href@noop {} {\bibfield  {journal} {\bibinfo  {journal}
  {Nature photonics}\ }\textbf {\bibinfo {volume} {11}},\ \bibinfo {pages}
  {441} (\bibinfo {year} {2017})}\BibitemShut {NoStop}%
\bibitem [{\citenamefont {Bueno}\ \emph {et~al.}(2018)\citenamefont {Bueno},
  \citenamefont {Maktoobi}, \citenamefont {Froehly}, \citenamefont {Fischer},
  \citenamefont {Jacquot}, \citenamefont {Larger},\ and\ \citenamefont
  {Brunner}}]{bueno2018reinforcement}%
  \BibitemOpen
  \bibfield  {author} {\bibinfo {author} {\bibfnamefont {J.}~\bibnamefont
  {Bueno}}, \bibinfo {author} {\bibfnamefont {S.}~\bibnamefont {Maktoobi}},
  \bibinfo {author} {\bibfnamefont {L.}~\bibnamefont {Froehly}}, \bibinfo
  {author} {\bibfnamefont {I.}~\bibnamefont {Fischer}}, \bibinfo {author}
  {\bibfnamefont {M.}~\bibnamefont {Jacquot}}, \bibinfo {author} {\bibfnamefont
  {L.}~\bibnamefont {Larger}}, \ and\ \bibinfo {author} {\bibfnamefont
  {D.}~\bibnamefont {Brunner}},\ }\href@noop {} {\bibfield  {journal} {\bibinfo
   {journal} {Optica}\ }\textbf {\bibinfo {volume} {5}},\ \bibinfo {pages}
  {756} (\bibinfo {year} {2018})}\BibitemShut {NoStop}%
\bibitem [{\citenamefont {Feldmann}\ \emph {et~al.}(2019)\citenamefont
  {Feldmann}, \citenamefont {Youngblood}, \citenamefont {Wright}, \citenamefont
  {Bhaskaran},\ and\ \citenamefont {Pernice}}]{feldmann2019all}%
  \BibitemOpen
  \bibfield  {author} {\bibinfo {author} {\bibfnamefont {J.}~\bibnamefont
  {Feldmann}}, \bibinfo {author} {\bibfnamefont {N.}~\bibnamefont
  {Youngblood}}, \bibinfo {author} {\bibfnamefont {C.~D.}\ \bibnamefont
  {Wright}}, \bibinfo {author} {\bibfnamefont {H.}~\bibnamefont {Bhaskaran}}, \
  and\ \bibinfo {author} {\bibfnamefont {W.~H.}\ \bibnamefont {Pernice}},\
  }\href@noop {} {\bibfield  {journal} {\bibinfo  {journal} {Nature}\ }\textbf
  {\bibinfo {volume} {569}},\ \bibinfo {pages} {208} (\bibinfo {year}
  {2019})}\BibitemShut {NoStop}%
\bibitem [{\citenamefont {Feldmann}\ \emph {et~al.}(2021)\citenamefont
  {Feldmann}, \citenamefont {Youngblood}, \citenamefont {Karpov}, \citenamefont
  {Gehring}, \citenamefont {Li}, \citenamefont {Stappers}, \citenamefont
  {Le~Gallo}, \citenamefont {Fu}, \citenamefont {Lukashchuk}, \citenamefont
  {Raja} \emph {et~al.}}]{feldmann2021parallel}%
  \BibitemOpen
  \bibfield  {author} {\bibinfo {author} {\bibfnamefont {J.}~\bibnamefont
  {Feldmann}}, \bibinfo {author} {\bibfnamefont {N.}~\bibnamefont
  {Youngblood}}, \bibinfo {author} {\bibfnamefont {M.}~\bibnamefont {Karpov}},
  \bibinfo {author} {\bibfnamefont {H.}~\bibnamefont {Gehring}}, \bibinfo
  {author} {\bibfnamefont {X.}~\bibnamefont {Li}}, \bibinfo {author}
  {\bibfnamefont {M.}~\bibnamefont {Stappers}}, \bibinfo {author}
  {\bibfnamefont {M.}~\bibnamefont {Le~Gallo}}, \bibinfo {author}
  {\bibfnamefont {X.}~\bibnamefont {Fu}}, \bibinfo {author} {\bibfnamefont
  {A.}~\bibnamefont {Lukashchuk}}, \bibinfo {author} {\bibfnamefont {A.~S.}\
  \bibnamefont {Raja}},  \emph {et~al.},\ }\href@noop {} {\bibfield  {journal}
  {\bibinfo  {journal} {Nature}\ }\textbf {\bibinfo {volume} {589}},\ \bibinfo
  {pages} {52} (\bibinfo {year} {2021})}\BibitemShut {NoStop}%
\bibitem [{\citenamefont {Pai}\ \emph {et~al.}(2023)\citenamefont {Pai},
  \citenamefont {Sun}, \citenamefont {Hughes}, \citenamefont {Park},
  \citenamefont {Bartlett}, \citenamefont {Williamson}, \citenamefont {Minkov},
  \citenamefont {Milanizadeh}, \citenamefont {Abebe}, \citenamefont
  {Morichetti} \emph {et~al.}}]{pai2023experimentally}%
  \BibitemOpen
  \bibfield  {author} {\bibinfo {author} {\bibfnamefont {S.}~\bibnamefont
  {Pai}}, \bibinfo {author} {\bibfnamefont {Z.}~\bibnamefont {Sun}}, \bibinfo
  {author} {\bibfnamefont {T.~W.}\ \bibnamefont {Hughes}}, \bibinfo {author}
  {\bibfnamefont {T.}~\bibnamefont {Park}}, \bibinfo {author} {\bibfnamefont
  {B.}~\bibnamefont {Bartlett}}, \bibinfo {author} {\bibfnamefont {I.~A.}\
  \bibnamefont {Williamson}}, \bibinfo {author} {\bibfnamefont
  {M.}~\bibnamefont {Minkov}}, \bibinfo {author} {\bibfnamefont
  {M.}~\bibnamefont {Milanizadeh}}, \bibinfo {author} {\bibfnamefont
  {N.}~\bibnamefont {Abebe}}, \bibinfo {author} {\bibfnamefont
  {F.}~\bibnamefont {Morichetti}},  \emph {et~al.},\ }\href@noop {} {\bibfield
  {journal} {\bibinfo  {journal} {Science}\ }\textbf {\bibinfo {volume}
  {380}},\ \bibinfo {pages} {398} (\bibinfo {year} {2023})}\BibitemShut
  {NoStop}%
\bibitem [{\citenamefont {Pashine}(2021)}]{pashine2021local}%
  \BibitemOpen
  \bibfield  {author} {\bibinfo {author} {\bibfnamefont {N.}~\bibnamefont
  {Pashine}},\ }\href@noop {} {\bibfield  {journal} {\bibinfo  {journal}
  {Physical Review Materials}\ }\textbf {\bibinfo {volume} {5}},\ \bibinfo
  {pages} {065607} (\bibinfo {year} {2021})}\BibitemShut {NoStop}%
\bibitem [{\citenamefont {Falk}\ \emph
  {et~al.}(2023{\natexlab{a}})\citenamefont {Falk}, \citenamefont {Wu},
  \citenamefont {Matthews}, \citenamefont {Sachdeva}, \citenamefont {Pashine},
  \citenamefont {Gardel}, \citenamefont {Nagel},\ and\ \citenamefont
  {Murugan}}]{falk2023learning}%
  \BibitemOpen
  \bibfield  {author} {\bibinfo {author} {\bibfnamefont {M.~J.}\ \bibnamefont
  {Falk}}, \bibinfo {author} {\bibfnamefont {J.}~\bibnamefont {Wu}}, \bibinfo
  {author} {\bibfnamefont {A.}~\bibnamefont {Matthews}}, \bibinfo {author}
  {\bibfnamefont {V.}~\bibnamefont {Sachdeva}}, \bibinfo {author}
  {\bibfnamefont {N.}~\bibnamefont {Pashine}}, \bibinfo {author} {\bibfnamefont
  {M.~L.}\ \bibnamefont {Gardel}}, \bibinfo {author} {\bibfnamefont {S.~R.}\
  \bibnamefont {Nagel}}, \ and\ \bibinfo {author} {\bibfnamefont
  {A.}~\bibnamefont {Murugan}},\ }\href@noop {} {\bibfield  {journal} {\bibinfo
   {journal} {Proceedings of the National Academy of Sciences}\ }\textbf
  {\bibinfo {volume} {120}},\ \bibinfo {pages} {e2219558120} (\bibinfo {year}
  {2023}{\natexlab{a}})}\BibitemShut {NoStop}%
\bibitem [{\citenamefont {Stern}\ and\ \citenamefont
  {Murugan}(2023)}]{stern2023learning}%
  \BibitemOpen
  \bibfield  {author} {\bibinfo {author} {\bibfnamefont {M.}~\bibnamefont
  {Stern}}\ and\ \bibinfo {author} {\bibfnamefont {A.}~\bibnamefont
  {Murugan}},\ }\href@noop {} {\bibfield  {journal} {\bibinfo  {journal}
  {Annual Review of Condensed Matter Physics}\ }\textbf {\bibinfo {volume}
  {14}},\ \bibinfo {pages} {417} (\bibinfo {year} {2023})}\BibitemShut
  {NoStop}%
\bibitem [{\citenamefont {Altman}\ \emph {et~al.}(2023)\citenamefont {Altman},
  \citenamefont {Stern}, \citenamefont {Liu},\ and\ \citenamefont
  {Durian}}]{altman2023experimental}%
  \BibitemOpen
  \bibfield  {author} {\bibinfo {author} {\bibfnamefont {L.~E.}\ \bibnamefont
  {Altman}}, \bibinfo {author} {\bibfnamefont {M.}~\bibnamefont {Stern}},
  \bibinfo {author} {\bibfnamefont {A.~J.}\ \bibnamefont {Liu}}, \ and\
  \bibinfo {author} {\bibfnamefont {D.~J.}\ \bibnamefont {Durian}},\
  }\href@noop {} {\bibfield  {journal} {\bibinfo  {journal} {arXiv preprint
  arXiv:2311.00170}\ } (\bibinfo {year} {2023})}\BibitemShut {NoStop}%
\bibitem [{\citenamefont {Acebr\'on}\ \emph {et~al.}(2005)\citenamefont
  {Acebr\'on}, \citenamefont {Bonilla}, \citenamefont {P\'erez~Vicente},
  \citenamefont {Ritort},\ and\ \citenamefont
  {Spigler}}]{KuramotoReview_Acebron}%
  \BibitemOpen
  \bibfield  {author} {\bibinfo {author} {\bibfnamefont {J.~A.}\ \bibnamefont
  {Acebr\'on}}, \bibinfo {author} {\bibfnamefont {L.~L.}\ \bibnamefont
  {Bonilla}}, \bibinfo {author} {\bibfnamefont {C.~J.}\ \bibnamefont
  {P\'erez~Vicente}}, \bibinfo {author} {\bibfnamefont {F.}~\bibnamefont
  {Ritort}}, \ and\ \bibinfo {author} {\bibfnamefont {R.}~\bibnamefont
  {Spigler}},\ }\href {\doibase 10.1103/RevModPhys.77.137} {\bibfield
  {journal} {\bibinfo  {journal} {Rev. Mod. Phys.}\ }\textbf {\bibinfo {volume}
  {77}},\ \bibinfo {pages} {137} (\bibinfo {year} {2005})}\BibitemShut
  {NoStop}%
\bibitem [{\citenamefont {Kosterlitz}(1974)}]{kosterlitz1974critical}%
  \BibitemOpen
  \bibfield  {author} {\bibinfo {author} {\bibfnamefont {J.}~\bibnamefont
  {Kosterlitz}},\ }\href@noop {} {\bibfield  {journal} {\bibinfo  {journal}
  {Journal of Physics C: Solid State Physics}\ }\textbf {\bibinfo {volume}
  {7}},\ \bibinfo {pages} {1046} (\bibinfo {year} {1974})}\BibitemShut
  {NoStop}%
\bibitem [{\citenamefont {Nixon}\ \emph {et~al.}(2013)\citenamefont {Nixon},
  \citenamefont {Ronen}, \citenamefont {Friesem},\ and\ \citenamefont
  {Davidson}}]{nixon2013observing}%
  \BibitemOpen
  \bibfield  {author} {\bibinfo {author} {\bibfnamefont {M.}~\bibnamefont
  {Nixon}}, \bibinfo {author} {\bibfnamefont {E.}~\bibnamefont {Ronen}},
  \bibinfo {author} {\bibfnamefont {A.~A.}\ \bibnamefont {Friesem}}, \ and\
  \bibinfo {author} {\bibfnamefont {N.}~\bibnamefont {Davidson}},\ }\href@noop
  {} {\bibfield  {journal} {\bibinfo  {journal} {Physical review letters}\
  }\textbf {\bibinfo {volume} {110}},\ \bibinfo {pages} {184102} (\bibinfo
  {year} {2013})}\BibitemShut {NoStop}%
\bibitem [{\citenamefont {Takeda}\ \emph {et~al.}(2017)\citenamefont {Takeda},
  \citenamefont {Tamate}, \citenamefont {Yamamoto}, \citenamefont {Takesue},
  \citenamefont {Inagaki},\ and\ \citenamefont
  {Utsunomiya}}]{takeda2017boltzmann}%
  \BibitemOpen
  \bibfield  {author} {\bibinfo {author} {\bibfnamefont {Y.}~\bibnamefont
  {Takeda}}, \bibinfo {author} {\bibfnamefont {S.}~\bibnamefont {Tamate}},
  \bibinfo {author} {\bibfnamefont {Y.}~\bibnamefont {Yamamoto}}, \bibinfo
  {author} {\bibfnamefont {H.}~\bibnamefont {Takesue}}, \bibinfo {author}
  {\bibfnamefont {T.}~\bibnamefont {Inagaki}}, \ and\ \bibinfo {author}
  {\bibfnamefont {S.}~\bibnamefont {Utsunomiya}},\ }\href@noop {} {\bibfield
  {journal} {\bibinfo  {journal} {Quantum Science and Technology}\ }\textbf
  {\bibinfo {volume} {3}},\ \bibinfo {pages} {014004} (\bibinfo {year}
  {2017})}\BibitemShut {NoStop}%
\bibitem [{\citenamefont {Heinrich}\ \emph {et~al.}(2011)\citenamefont
  {Heinrich}, \citenamefont {Ludwig}, \citenamefont {Qian}, \citenamefont
  {Kubala},\ and\ \citenamefont {Marquardt}}]{Heinrich2011Collective}%
  \BibitemOpen
  \bibfield  {author} {\bibinfo {author} {\bibfnamefont {G.}~\bibnamefont
  {Heinrich}}, \bibinfo {author} {\bibfnamefont {M.}~\bibnamefont {Ludwig}},
  \bibinfo {author} {\bibfnamefont {J.}~\bibnamefont {Qian}}, \bibinfo {author}
  {\bibfnamefont {B.}~\bibnamefont {Kubala}}, \ and\ \bibinfo {author}
  {\bibfnamefont {F.}~\bibnamefont {Marquardt}},\ }\href {\doibase
  10.1103/PhysRevLett.107.043603} {\bibfield  {journal} {\bibinfo  {journal}
  {Phys. Rev. Lett.}\ }\textbf {\bibinfo {volume} {107}},\ \bibinfo {pages}
  {043603} (\bibinfo {year} {2011})}\BibitemShut {NoStop}%
\bibitem [{\citenamefont {Zhang}\ \emph {et~al.}(2015)\citenamefont {Zhang},
  \citenamefont {Shah}, \citenamefont {Cardenas},\ and\ \citenamefont
  {Lipson}}]{zhang2015synchronization}%
  \BibitemOpen
  \bibfield  {author} {\bibinfo {author} {\bibfnamefont {M.}~\bibnamefont
  {Zhang}}, \bibinfo {author} {\bibfnamefont {S.}~\bibnamefont {Shah}},
  \bibinfo {author} {\bibfnamefont {J.}~\bibnamefont {Cardenas}}, \ and\
  \bibinfo {author} {\bibfnamefont {M.}~\bibnamefont {Lipson}},\ }\href@noop {}
  {\bibfield  {journal} {\bibinfo  {journal} {Physical review letters}\
  }\textbf {\bibinfo {volume} {115}},\ \bibinfo {pages} {163902} (\bibinfo
  {year} {2015})}\BibitemShut {NoStop}%
\bibitem [{\citenamefont {Matheny}\ \emph {et~al.}(2014)\citenamefont
  {Matheny}, \citenamefont {Grau}, \citenamefont {Villanueva}, \citenamefont
  {Karabalin}, \citenamefont {Cross},\ and\ \citenamefont
  {Roukes}}]{matheny2014phase}%
  \BibitemOpen
  \bibfield  {author} {\bibinfo {author} {\bibfnamefont {M.~H.}\ \bibnamefont
  {Matheny}}, \bibinfo {author} {\bibfnamefont {M.}~\bibnamefont {Grau}},
  \bibinfo {author} {\bibfnamefont {L.~G.}\ \bibnamefont {Villanueva}},
  \bibinfo {author} {\bibfnamefont {R.~B.}\ \bibnamefont {Karabalin}}, \bibinfo
  {author} {\bibfnamefont {M.}~\bibnamefont {Cross}}, \ and\ \bibinfo {author}
  {\bibfnamefont {M.~L.}\ \bibnamefont {Roukes}},\ }\href@noop {} {\bibfield
  {journal} {\bibinfo  {journal} {Physical review letters}\ }\textbf {\bibinfo
  {volume} {112}},\ \bibinfo {pages} {014101} (\bibinfo {year}
  {2014})}\BibitemShut {NoStop}%
\bibitem [{\citenamefont {Struck}\ \emph {et~al.}(2013)\citenamefont {Struck},
  \citenamefont {Weinberg}, \citenamefont {{\"O}lschl{\"a}ger}, \citenamefont
  {Windpassinger}, \citenamefont {Simonet}, \citenamefont {Sengstock},
  \citenamefont {H{\"o}ppner}, \citenamefont {Hauke}, \citenamefont {Eckardt},
  \citenamefont {Lewenstein} \emph {et~al.}}]{struck2013engineering}%
  \BibitemOpen
  \bibfield  {author} {\bibinfo {author} {\bibfnamefont {J.}~\bibnamefont
  {Struck}}, \bibinfo {author} {\bibfnamefont {M.}~\bibnamefont {Weinberg}},
  \bibinfo {author} {\bibfnamefont {C.}~\bibnamefont {{\"O}lschl{\"a}ger}},
  \bibinfo {author} {\bibfnamefont {P.}~\bibnamefont {Windpassinger}}, \bibinfo
  {author} {\bibfnamefont {J.}~\bibnamefont {Simonet}}, \bibinfo {author}
  {\bibfnamefont {K.}~\bibnamefont {Sengstock}}, \bibinfo {author}
  {\bibfnamefont {R.}~\bibnamefont {H{\"o}ppner}}, \bibinfo {author}
  {\bibfnamefont {P.}~\bibnamefont {Hauke}}, \bibinfo {author} {\bibfnamefont
  {A.}~\bibnamefont {Eckardt}}, \bibinfo {author} {\bibfnamefont
  {M.}~\bibnamefont {Lewenstein}},  \emph {et~al.},\ }\href@noop {} {\bibfield
  {journal} {\bibinfo  {journal} {Nature Physics}\ }\textbf {\bibinfo {volume}
  {9}},\ \bibinfo {pages} {738} (\bibinfo {year} {2013})}\BibitemShut {NoStop}%
\bibitem [{\citenamefont {Cosmic}\ \emph {et~al.}(2020)\citenamefont {Cosmic},
  \citenamefont {Kawabata}, \citenamefont {Ashida}, \citenamefont {Ikegami},
  \citenamefont {Furukawa}, \citenamefont {Patil}, \citenamefont {Taylor},\
  and\ \citenamefont {Nakamura}}]{Cosmic2020Probing}%
  \BibitemOpen
  \bibfield  {author} {\bibinfo {author} {\bibfnamefont {R.}~\bibnamefont
  {Cosmic}}, \bibinfo {author} {\bibfnamefont {K.}~\bibnamefont {Kawabata}},
  \bibinfo {author} {\bibfnamefont {Y.}~\bibnamefont {Ashida}}, \bibinfo
  {author} {\bibfnamefont {H.}~\bibnamefont {Ikegami}}, \bibinfo {author}
  {\bibfnamefont {S.}~\bibnamefont {Furukawa}}, \bibinfo {author}
  {\bibfnamefont {P.}~\bibnamefont {Patil}}, \bibinfo {author} {\bibfnamefont
  {J.~M.}\ \bibnamefont {Taylor}}, \ and\ \bibinfo {author} {\bibfnamefont
  {Y.}~\bibnamefont {Nakamura}},\ }\href {\doibase 10.1103/PhysRevB.102.094509}
  {\bibfield  {journal} {\bibinfo  {journal} {Phys. Rev. B}\ }\textbf {\bibinfo
  {volume} {102}},\ \bibinfo {pages} {094509} (\bibinfo {year}
  {2020})}\BibitemShut {NoStop}%
\bibitem [{\citenamefont {Torrejon}\ \emph
  {et~al.}(2017{\natexlab{b}})\citenamefont {Torrejon}, \citenamefont {Riou},
  \citenamefont {Araujo}, \citenamefont {Tsunegi}, \citenamefont {Khalsa},
  \citenamefont {Querlioz}, \citenamefont {Bortolotti}, \citenamefont {Cros},
  \citenamefont {Yakushiji}, \citenamefont {Fukushima}, \citenamefont {Kubota},
  \citenamefont {Yuasa}, \citenamefont {Stiles},\ and\ \citenamefont
  {Grollier}}]{Torrejon2017Neuromorphic}%
  \BibitemOpen
  \bibfield  {author} {\bibinfo {author} {\bibfnamefont {J.}~\bibnamefont
  {Torrejon}}, \bibinfo {author} {\bibfnamefont {M.}~\bibnamefont {Riou}},
  \bibinfo {author} {\bibfnamefont {F.~A.}\ \bibnamefont {Araujo}}, \bibinfo
  {author} {\bibfnamefont {S.}~\bibnamefont {Tsunegi}}, \bibinfo {author}
  {\bibfnamefont {G.}~\bibnamefont {Khalsa}}, \bibinfo {author} {\bibfnamefont
  {D.}~\bibnamefont {Querlioz}}, \bibinfo {author} {\bibfnamefont
  {P.}~\bibnamefont {Bortolotti}}, \bibinfo {author} {\bibfnamefont
  {V.}~\bibnamefont {Cros}}, \bibinfo {author} {\bibfnamefont {K.}~\bibnamefont
  {Yakushiji}}, \bibinfo {author} {\bibfnamefont {A.}~\bibnamefont
  {Fukushima}}, \bibinfo {author} {\bibfnamefont {H.}~\bibnamefont {Kubota}},
  \bibinfo {author} {\bibfnamefont {S.}~\bibnamefont {Yuasa}}, \bibinfo
  {author} {\bibfnamefont {M.~D.}\ \bibnamefont {Stiles}}, \ and\ \bibinfo
  {author} {\bibfnamefont {J.}~\bibnamefont {Grollier}},\ }\href {\doibase
  10.1038/nature23011} {\bibfield  {journal} {\bibinfo  {journal} {Nature}\
  }\textbf {\bibinfo {volume} {547}},\ \bibinfo {pages} {428} (\bibinfo {year}
  {2017}{\natexlab{b}})}\BibitemShut {NoStop}%
\bibitem [{\citenamefont {Romera}\ \emph {et~al.}(2018)\citenamefont {Romera},
  \citenamefont {Talatchian}, \citenamefont {Tsunegi}, \citenamefont
  {Abreu~Araujo}, \citenamefont {Cros}, \citenamefont {Bortolotti},
  \citenamefont {Trastoy}, \citenamefont {Yakushiji}, \citenamefont
  {Fukushima}, \citenamefont {Kubota} \emph {et~al.}}]{romera2018vowel}%
  \BibitemOpen
  \bibfield  {author} {\bibinfo {author} {\bibfnamefont {M.}~\bibnamefont
  {Romera}}, \bibinfo {author} {\bibfnamefont {P.}~\bibnamefont {Talatchian}},
  \bibinfo {author} {\bibfnamefont {S.}~\bibnamefont {Tsunegi}}, \bibinfo
  {author} {\bibfnamefont {F.}~\bibnamefont {Abreu~Araujo}}, \bibinfo {author}
  {\bibfnamefont {V.}~\bibnamefont {Cros}}, \bibinfo {author} {\bibfnamefont
  {P.}~\bibnamefont {Bortolotti}}, \bibinfo {author} {\bibfnamefont
  {J.}~\bibnamefont {Trastoy}}, \bibinfo {author} {\bibfnamefont
  {K.}~\bibnamefont {Yakushiji}}, \bibinfo {author} {\bibfnamefont
  {A.}~\bibnamefont {Fukushima}}, \bibinfo {author} {\bibfnamefont
  {H.}~\bibnamefont {Kubota}},  \emph {et~al.},\ }\href@noop {} {\bibfield
  {journal} {\bibinfo  {journal} {Nature}\ }\textbf {\bibinfo {volume} {563}},\
  \bibinfo {pages} {230} (\bibinfo {year} {2018})}\BibitemShut {NoStop}%
\bibitem [{\citenamefont {Baas}\ \emph {et~al.}(2008)\citenamefont {Baas},
  \citenamefont {Lagoudakis}, \citenamefont {Richard}, \citenamefont
  {Andr{\'e}}, \citenamefont {Dang},\ and\ \citenamefont
  {Deveaud-Pl{\'e}dran}}]{baas2008synchronized}%
  \BibitemOpen
  \bibfield  {author} {\bibinfo {author} {\bibfnamefont {A.}~\bibnamefont
  {Baas}}, \bibinfo {author} {\bibfnamefont {K.}~\bibnamefont {Lagoudakis}},
  \bibinfo {author} {\bibfnamefont {M.}~\bibnamefont {Richard}}, \bibinfo
  {author} {\bibfnamefont {R.}~\bibnamefont {Andr{\'e}}}, \bibinfo {author}
  {\bibfnamefont {L.~S.}\ \bibnamefont {Dang}}, \ and\ \bibinfo {author}
  {\bibfnamefont {B.}~\bibnamefont {Deveaud-Pl{\'e}dran}},\ }\href@noop {}
  {\bibfield  {journal} {\bibinfo  {journal} {Physical review letters}\
  }\textbf {\bibinfo {volume} {100}},\ \bibinfo {pages} {170401} (\bibinfo
  {year} {2008})}\BibitemShut {NoStop}%
\bibitem [{\citenamefont {Berloff}\ \emph {et~al.}(2017)\citenamefont
  {Berloff}, \citenamefont {Silva}, \citenamefont {Kalinin}, \citenamefont
  {Askitopoulos}, \citenamefont {T{\"o}pfer}, \citenamefont {Cilibrizzi},
  \citenamefont {Langbein},\ and\ \citenamefont
  {Lagoudakis}}]{Berloff2017Realizing}%
  \BibitemOpen
  \bibfield  {author} {\bibinfo {author} {\bibfnamefont {N.~G.}\ \bibnamefont
  {Berloff}}, \bibinfo {author} {\bibfnamefont {M.}~\bibnamefont {Silva}},
  \bibinfo {author} {\bibfnamefont {K.}~\bibnamefont {Kalinin}}, \bibinfo
  {author} {\bibfnamefont {A.}~\bibnamefont {Askitopoulos}}, \bibinfo {author}
  {\bibfnamefont {J.~D.}\ \bibnamefont {T{\"o}pfer}}, \bibinfo {author}
  {\bibfnamefont {P.}~\bibnamefont {Cilibrizzi}}, \bibinfo {author}
  {\bibfnamefont {W.}~\bibnamefont {Langbein}}, \ and\ \bibinfo {author}
  {\bibfnamefont {P.~G.}\ \bibnamefont {Lagoudakis}},\ }\href {\doibase
  10.1038/nmat4971} {\bibfield  {journal} {\bibinfo  {journal} {Nature
  Materials}\ }\textbf {\bibinfo {volume} {16}},\ \bibinfo {pages} {1120}
  (\bibinfo {year} {2017})}\BibitemShut {NoStop}%
\bibitem [{\citenamefont {Kavokin}\ \emph {et~al.}(2022)\citenamefont
  {Kavokin}, \citenamefont {Liew}, \citenamefont {Schneider}, \citenamefont
  {Lagoudakis}, \citenamefont {Klembt},\ and\ \citenamefont
  {Hoefling}}]{Kavokin2022Polariton}%
  \BibitemOpen
  \bibfield  {author} {\bibinfo {author} {\bibfnamefont {A.}~\bibnamefont
  {Kavokin}}, \bibinfo {author} {\bibfnamefont {T.~C.~H.}\ \bibnamefont
  {Liew}}, \bibinfo {author} {\bibfnamefont {C.}~\bibnamefont {Schneider}},
  \bibinfo {author} {\bibfnamefont {P.~G.}\ \bibnamefont {Lagoudakis}},
  \bibinfo {author} {\bibfnamefont {S.}~\bibnamefont {Klembt}}, \ and\ \bibinfo
  {author} {\bibfnamefont {S.}~\bibnamefont {Hoefling}},\ }\href {\doibase
  10.1038/s42254-022-00447-1} {\bibfield  {journal} {\bibinfo  {journal}
  {Nature Reviews Physics}\ }\textbf {\bibinfo {volume} {4}},\ \bibinfo {pages}
  {435} (\bibinfo {year} {2022})}\BibitemShut {NoStop}%
\bibitem [{\citenamefont {Hoppensteadt}\ and\ \citenamefont
  {Izhikevich}(1999)}]{hoppensteadt1999oscillatory}%
  \BibitemOpen
  \bibfield  {author} {\bibinfo {author} {\bibfnamefont {F.~C.}\ \bibnamefont
  {Hoppensteadt}}\ and\ \bibinfo {author} {\bibfnamefont {E.~M.}\ \bibnamefont
  {Izhikevich}},\ }\href@noop {} {\bibfield  {journal} {\bibinfo  {journal}
  {Physical Review Letters}\ }\textbf {\bibinfo {volume} {82}},\ \bibinfo
  {pages} {2983} (\bibinfo {year} {1999})}\BibitemShut {NoStop}%
\bibitem [{\citenamefont {Hoppensteadt}\ and\ \citenamefont
  {Izhikevich}(2000)}]{hoppensteadt2000synchronization}%
  \BibitemOpen
  \bibfield  {author} {\bibinfo {author} {\bibfnamefont {F.~C.}\ \bibnamefont
  {Hoppensteadt}}\ and\ \bibinfo {author} {\bibfnamefont {E.~M.}\ \bibnamefont
  {Izhikevich}},\ }\href@noop {} {\bibfield  {journal} {\bibinfo  {journal}
  {Physical Review E}\ }\textbf {\bibinfo {volume} {62}},\ \bibinfo {pages}
  {4010} (\bibinfo {year} {2000})}\BibitemShut {NoStop}%
\bibitem [{\citenamefont {Stroev}\ and\ \citenamefont
  {Berloff}(2021)}]{stroev2021neural}%
  \BibitemOpen
  \bibfield  {author} {\bibinfo {author} {\bibfnamefont {N.}~\bibnamefont
  {Stroev}}\ and\ \bibinfo {author} {\bibfnamefont {N.~G.}\ \bibnamefont
  {Berloff}},\ }\href@noop {} {\bibfield  {journal} {\bibinfo  {journal}
  {Physical Review B}\ }\textbf {\bibinfo {volume} {104}},\ \bibinfo {pages}
  {205435} (\bibinfo {year} {2021})}\BibitemShut {NoStop}%
\bibitem [{\citenamefont {Csaba}\ and\ \citenamefont
  {Porod}(2020)}]{csaba2020coupled}%
  \BibitemOpen
  \bibfield  {author} {\bibinfo {author} {\bibfnamefont {G.}~\bibnamefont
  {Csaba}}\ and\ \bibinfo {author} {\bibfnamefont {W.}~\bibnamefont {Porod}},\
  }\href@noop {} {\bibfield  {journal} {\bibinfo  {journal} {Applied physics
  reviews}\ }\textbf {\bibinfo {volume} {7}} (\bibinfo {year}
  {2020})}\BibitemShut {NoStop}%
\bibitem [{\citenamefont {Rudner}\ \emph {et~al.}(2023)\citenamefont {Rudner},
  \citenamefont {Porod},\ and\ \citenamefont {Csaba}}]{rudner2023design}%
  \BibitemOpen
  \bibfield  {author} {\bibinfo {author} {\bibfnamefont {T.}~\bibnamefont
  {Rudner}}, \bibinfo {author} {\bibfnamefont {W.}~\bibnamefont {Porod}}, \
  and\ \bibinfo {author} {\bibfnamefont {G.}~\bibnamefont {Csaba}},\
  }\href@noop {} {\bibfield  {journal} {\bibinfo  {journal} {arXiv preprint
  arXiv:2309.02532}\ } (\bibinfo {year} {2023})}\BibitemShut {NoStop}%
\bibitem [{\citenamefont {Filipovich}\ \emph {et~al.}(2022)\citenamefont
  {Filipovich}, \citenamefont {Guo}, \citenamefont {Al-Qadasi}, \citenamefont
  {Marquez}, \citenamefont {Morison}, \citenamefont {Sorger}, \citenamefont
  {Prucnal}, \citenamefont {Shekhar},\ and\ \citenamefont
  {Shastri}}]{Filipovich2022Silicon}%
  \BibitemOpen
  \bibfield  {author} {\bibinfo {author} {\bibfnamefont {M.~J.}\ \bibnamefont
  {Filipovich}}, \bibinfo {author} {\bibfnamefont {Z.}~\bibnamefont {Guo}},
  \bibinfo {author} {\bibfnamefont {M.}~\bibnamefont {Al-Qadasi}}, \bibinfo
  {author} {\bibfnamefont {B.~A.}\ \bibnamefont {Marquez}}, \bibinfo {author}
  {\bibfnamefont {H.~D.}\ \bibnamefont {Morison}}, \bibinfo {author}
  {\bibfnamefont {V.~J.}\ \bibnamefont {Sorger}}, \bibinfo {author}
  {\bibfnamefont {P.~R.}\ \bibnamefont {Prucnal}}, \bibinfo {author}
  {\bibfnamefont {S.}~\bibnamefont {Shekhar}}, \ and\ \bibinfo {author}
  {\bibfnamefont {B.~J.}\ \bibnamefont {Shastri}},\ }\href {\doibase
  10.1364/OPTICA.475493} {\bibfield  {journal} {\bibinfo  {journal} {Optica}\
  }\textbf {\bibinfo {volume} {9}},\ \bibinfo {pages} {1323} (\bibinfo {year}
  {2022})}\BibitemShut {NoStop}%
\bibitem [{\citenamefont {Bandyopadhyay}\ \emph {et~al.}(2022)\citenamefont
  {Bandyopadhyay}, \citenamefont {Sludds}, \citenamefont {Krastanov},
  \citenamefont {Hamerly}, \citenamefont {Harris}, \citenamefont {Bunandar},
  \citenamefont {Streshinsky}, \citenamefont {Hochberg},\ and\ \citenamefont
  {Englund}}]{bandyopadhyay2022single}%
  \BibitemOpen
  \bibfield  {author} {\bibinfo {author} {\bibfnamefont {S.}~\bibnamefont
  {Bandyopadhyay}}, \bibinfo {author} {\bibfnamefont {A.}~\bibnamefont
  {Sludds}}, \bibinfo {author} {\bibfnamefont {S.}~\bibnamefont {Krastanov}},
  \bibinfo {author} {\bibfnamefont {R.}~\bibnamefont {Hamerly}}, \bibinfo
  {author} {\bibfnamefont {N.}~\bibnamefont {Harris}}, \bibinfo {author}
  {\bibfnamefont {D.}~\bibnamefont {Bunandar}}, \bibinfo {author}
  {\bibfnamefont {M.}~\bibnamefont {Streshinsky}}, \bibinfo {author}
  {\bibfnamefont {M.}~\bibnamefont {Hochberg}}, \ and\ \bibinfo {author}
  {\bibfnamefont {D.}~\bibnamefont {Englund}},\ }\href@noop {} {\  (\bibinfo
  {year} {2022})},\ \Eprint {http://arxiv.org/abs/2208.01623} {arXiv:2208.01623
  [cs.ET]} \BibitemShut {NoStop}%
\bibitem [{\citenamefont {Duport}\ \emph {et~al.}(2012)\citenamefont {Duport},
  \citenamefont {Schneider}, \citenamefont {Smerieri}, \citenamefont
  {Haelterman},\ and\ \citenamefont {Massar}}]{duport2012all}%
  \BibitemOpen
  \bibfield  {author} {\bibinfo {author} {\bibfnamefont {F.}~\bibnamefont
  {Duport}}, \bibinfo {author} {\bibfnamefont {B.}~\bibnamefont {Schneider}},
  \bibinfo {author} {\bibfnamefont {A.}~\bibnamefont {Smerieri}}, \bibinfo
  {author} {\bibfnamefont {M.}~\bibnamefont {Haelterman}}, \ and\ \bibinfo
  {author} {\bibfnamefont {S.}~\bibnamefont {Massar}},\ }\href@noop {}
  {\bibfield  {journal} {\bibinfo  {journal} {Optics express}\ }\textbf
  {\bibinfo {volume} {20}},\ \bibinfo {pages} {22783} (\bibinfo {year}
  {2012})}\BibitemShut {NoStop}%
\bibitem [{\citenamefont {Tanaka}\ \emph {et~al.}(2019)\citenamefont {Tanaka},
  \citenamefont {Yamane}, \citenamefont {H{\'e}roux}, \citenamefont {Nakane},
  \citenamefont {Kanazawa}, \citenamefont {Takeda}, \citenamefont {Numata},
  \citenamefont {Nakano},\ and\ \citenamefont {Hirose}}]{tanaka2019recent}%
  \BibitemOpen
  \bibfield  {author} {\bibinfo {author} {\bibfnamefont {G.}~\bibnamefont
  {Tanaka}}, \bibinfo {author} {\bibfnamefont {T.}~\bibnamefont {Yamane}},
  \bibinfo {author} {\bibfnamefont {J.~B.}\ \bibnamefont {H{\'e}roux}},
  \bibinfo {author} {\bibfnamefont {R.}~\bibnamefont {Nakane}}, \bibinfo
  {author} {\bibfnamefont {N.}~\bibnamefont {Kanazawa}}, \bibinfo {author}
  {\bibfnamefont {S.}~\bibnamefont {Takeda}}, \bibinfo {author} {\bibfnamefont
  {H.}~\bibnamefont {Numata}}, \bibinfo {author} {\bibfnamefont
  {D.}~\bibnamefont {Nakano}}, \ and\ \bibinfo {author} {\bibfnamefont
  {A.}~\bibnamefont {Hirose}},\ }\href@noop {} {\bibfield  {journal} {\bibinfo
  {journal} {Neural Networks}\ }\textbf {\bibinfo {volume} {115}},\ \bibinfo
  {pages} {100} (\bibinfo {year} {2019})}\BibitemShut {NoStop}%
\bibitem [{\citenamefont {Nakajima}(2020)}]{nakajima2020physical}%
  \BibitemOpen
  \bibfield  {author} {\bibinfo {author} {\bibfnamefont {K.}~\bibnamefont
  {Nakajima}},\ }\href@noop {} {\bibfield  {journal} {\bibinfo  {journal}
  {Japanese Journal of Applied Physics}\ }\textbf {\bibinfo {volume} {59}},\
  \bibinfo {pages} {060501} (\bibinfo {year} {2020})}\BibitemShut {NoStop}%
\bibitem [{\citenamefont {Van~der Sande}\ \emph {et~al.}(2017)\citenamefont
  {Van~der Sande}, \citenamefont {Brunner},\ and\ \citenamefont
  {Soriano}}]{van2017advances}%
  \BibitemOpen
  \bibfield  {author} {\bibinfo {author} {\bibfnamefont {G.}~\bibnamefont
  {Van~der Sande}}, \bibinfo {author} {\bibfnamefont {D.}~\bibnamefont
  {Brunner}}, \ and\ \bibinfo {author} {\bibfnamefont {M.~C.}\ \bibnamefont
  {Soriano}},\ }\href@noop {} {\bibfield  {journal} {\bibinfo  {journal}
  {Nanophotonics}\ }\textbf {\bibinfo {volume} {6}},\ \bibinfo {pages} {561}
  (\bibinfo {year} {2017})}\BibitemShut {NoStop}%
\bibitem [{\citenamefont {Wright}\ \emph {et~al.}(2022)\citenamefont {Wright},
  \citenamefont {Onodera}, \citenamefont {Stein}, \citenamefont {Wang},
  \citenamefont {Schachter}, \citenamefont {Hu},\ and\ \citenamefont
  {McMahon}}]{wright2022deep}%
  \BibitemOpen
  \bibfield  {author} {\bibinfo {author} {\bibfnamefont {L.~G.}\ \bibnamefont
  {Wright}}, \bibinfo {author} {\bibfnamefont {T.}~\bibnamefont {Onodera}},
  \bibinfo {author} {\bibfnamefont {M.~M.}\ \bibnamefont {Stein}}, \bibinfo
  {author} {\bibfnamefont {T.}~\bibnamefont {Wang}}, \bibinfo {author}
  {\bibfnamefont {D.~T.}\ \bibnamefont {Schachter}}, \bibinfo {author}
  {\bibfnamefont {Z.}~\bibnamefont {Hu}}, \ and\ \bibinfo {author}
  {\bibfnamefont {P.~L.}\ \bibnamefont {McMahon}},\ }\href@noop {} {\bibfield
  {journal} {\bibinfo  {journal} {Nature}\ }\textbf {\bibinfo {volume} {601}},\
  \bibinfo {pages} {549} (\bibinfo {year} {2022})}\BibitemShut {NoStop}%
\bibitem [{\citenamefont {Psaltis}\ \emph {et~al.}(1990)\citenamefont
  {Psaltis}, \citenamefont {Brady}, \citenamefont {Gu},\ and\ \citenamefont
  {Lin}}]{psaltis1990holography}%
  \BibitemOpen
  \bibfield  {author} {\bibinfo {author} {\bibfnamefont {D.}~\bibnamefont
  {Psaltis}}, \bibinfo {author} {\bibfnamefont {D.}~\bibnamefont {Brady}},
  \bibinfo {author} {\bibfnamefont {X.-G.}\ \bibnamefont {Gu}}, \ and\ \bibinfo
  {author} {\bibfnamefont {S.}~\bibnamefont {Lin}},\ }\href@noop {} {\bibfield
  {journal} {\bibinfo  {journal} {Nature}\ }\textbf {\bibinfo {volume} {343}},\
  \bibinfo {pages} {325} (\bibinfo {year} {1990})}\BibitemShut {NoStop}%
\bibitem [{\citenamefont {Guo}\ \emph {et~al.}(2021)\citenamefont {Guo},
  \citenamefont {Barrett}, \citenamefont {Wang},\ and\ \citenamefont
  {Lvovsky}}]{guo2021backpropagation}%
  \BibitemOpen
  \bibfield  {author} {\bibinfo {author} {\bibfnamefont {X.}~\bibnamefont
  {Guo}}, \bibinfo {author} {\bibfnamefont {T.~D.}\ \bibnamefont {Barrett}},
  \bibinfo {author} {\bibfnamefont {Z.~M.}\ \bibnamefont {Wang}}, \ and\
  \bibinfo {author} {\bibfnamefont {A.}~\bibnamefont {Lvovsky}},\ }\href@noop
  {} {\bibfield  {journal} {\bibinfo  {journal} {Photonics Research}\ }\textbf
  {\bibinfo {volume} {9}},\ \bibinfo {pages} {B71} (\bibinfo {year}
  {2021})}\BibitemShut {NoStop}%
\bibitem [{\citenamefont {Spall}\ \emph {et~al.}(2023)\citenamefont {Spall},
  \citenamefont {Guo},\ and\ \citenamefont {Lvovsky}}]{spall2023training}%
  \BibitemOpen
  \bibfield  {author} {\bibinfo {author} {\bibfnamefont {J.}~\bibnamefont
  {Spall}}, \bibinfo {author} {\bibfnamefont {X.}~\bibnamefont {Guo}}, \ and\
  \bibinfo {author} {\bibfnamefont {A.~I.}\ \bibnamefont {Lvovsky}},\
  }\href@noop {} {\enquote {\bibinfo {title} {Training neural networks with
  end-to-end optical backpropagation},}\ } (\bibinfo {year} {2023}),\ \Eprint
  {http://arxiv.org/abs/2308.05226} {arXiv:2308.05226 [physics.optics]}
  \BibitemShut {NoStop}%
\bibitem [{\citenamefont {Hughes}\ \emph {et~al.}(2018)\citenamefont {Hughes},
  \citenamefont {Minkov}, \citenamefont {Shi},\ and\ \citenamefont
  {Fan}}]{hughes2018training}%
  \BibitemOpen
  \bibfield  {author} {\bibinfo {author} {\bibfnamefont {T.~W.}\ \bibnamefont
  {Hughes}}, \bibinfo {author} {\bibfnamefont {M.}~\bibnamefont {Minkov}},
  \bibinfo {author} {\bibfnamefont {Y.}~\bibnamefont {Shi}}, \ and\ \bibinfo
  {author} {\bibfnamefont {S.}~\bibnamefont {Fan}},\ }\href@noop {} {\bibfield
  {journal} {\bibinfo  {journal} {Optica}\ }\textbf {\bibinfo {volume} {5}},\
  \bibinfo {pages} {864} (\bibinfo {year} {2018})}\BibitemShut {NoStop}%
\bibitem [{\citenamefont {Wanjura}\ and\ \citenamefont
  {Marquardt}(2023)}]{wanjura2023fully}%
  \BibitemOpen
  \bibfield  {author} {\bibinfo {author} {\bibfnamefont {C.~C.}\ \bibnamefont
  {Wanjura}}\ and\ \bibinfo {author} {\bibfnamefont {F.}~\bibnamefont
  {Marquardt}},\ }\href@noop {} {\bibfield  {journal} {\bibinfo  {journal}
  {arXiv preprint arXiv:2308.16181}\ } (\bibinfo {year} {2023})}\BibitemShut
  {NoStop}%
\bibitem [{\citenamefont {Lopez-Pastor}\ and\ \citenamefont
  {Marquardt}(2021)}]{lopez2021self}%
  \BibitemOpen
  \bibfield  {author} {\bibinfo {author} {\bibfnamefont {V.}~\bibnamefont
  {Lopez-Pastor}}\ and\ \bibinfo {author} {\bibfnamefont {F.}~\bibnamefont
  {Marquardt}},\ }\href@noop {} {\bibfield  {journal} {\bibinfo  {journal}
  {arXiv preprint arXiv:2103.04992}\ } (\bibinfo {year} {2021})}\BibitemShut
  {NoStop}%
\bibitem [{\citenamefont {Scellier}\ and\ \citenamefont
  {Bengio}(2017)}]{scellier2017equilibrium}%
  \BibitemOpen
  \bibfield  {author} {\bibinfo {author} {\bibfnamefont {B.}~\bibnamefont
  {Scellier}}\ and\ \bibinfo {author} {\bibfnamefont {Y.}~\bibnamefont
  {Bengio}},\ }\href@noop {} {\bibfield  {journal} {\bibinfo  {journal}
  {Frontiers in computational neuroscience}\ }\textbf {\bibinfo {volume}
  {11}},\ \bibinfo {pages} {24} (\bibinfo {year} {2017})}\BibitemShut {NoStop}%
\bibitem [{\citenamefont {Scellier}(2021)}]{scellier2021deep}%
  \BibitemOpen
  \bibfield  {author} {\bibinfo {author} {\bibfnamefont {B.}~\bibnamefont
  {Scellier}},\ }\href@noop {} {\bibfield  {journal} {\bibinfo  {journal}
  {arXiv preprint arXiv:2103.09985}\ } (\bibinfo {year} {2021})}\BibitemShut
  {NoStop}%
\bibitem [{\citenamefont {Ackley}\ \emph {et~al.}(1985)\citenamefont {Ackley},
  \citenamefont {Hinton},\ and\ \citenamefont
  {Sejnowski}}]{ackley1985learning}%
  \BibitemOpen
  \bibfield  {author} {\bibinfo {author} {\bibfnamefont {D.~H.}\ \bibnamefont
  {Ackley}}, \bibinfo {author} {\bibfnamefont {G.~E.}\ \bibnamefont {Hinton}},
  \ and\ \bibinfo {author} {\bibfnamefont {T.~J.}\ \bibnamefont {Sejnowski}},\
  }\href@noop {} {\bibfield  {journal} {\bibinfo  {journal} {Cognitive
  science}\ }\textbf {\bibinfo {volume} {9}},\ \bibinfo {pages} {147} (\bibinfo
  {year} {1985})}\BibitemShut {NoStop}%
\bibitem [{\citenamefont {Stern}\ \emph {et~al.}(2021)\citenamefont {Stern},
  \citenamefont {Hexner}, \citenamefont {Rocks},\ and\ \citenamefont
  {Liu}}]{stern2021supervised}%
  \BibitemOpen
  \bibfield  {author} {\bibinfo {author} {\bibfnamefont {M.}~\bibnamefont
  {Stern}}, \bibinfo {author} {\bibfnamefont {D.}~\bibnamefont {Hexner}},
  \bibinfo {author} {\bibfnamefont {J.~W.}\ \bibnamefont {Rocks}}, \ and\
  \bibinfo {author} {\bibfnamefont {A.~J.}\ \bibnamefont {Liu}},\ }\href@noop
  {} {\bibfield  {journal} {\bibinfo  {journal} {Physical Review X}\ }\textbf
  {\bibinfo {volume} {11}},\ \bibinfo {pages} {021045} (\bibinfo {year}
  {2021})}\BibitemShut {NoStop}%
\bibitem [{\citenamefont {Scellier}\ \emph {et~al.}(2023)\citenamefont
  {Scellier}, \citenamefont {Ernoult}, \citenamefont {Kendall},\ and\
  \citenamefont {Kumar}}]{scellier2023energybased}%
  \BibitemOpen
  \bibfield  {author} {\bibinfo {author} {\bibfnamefont {B.}~\bibnamefont
  {Scellier}}, \bibinfo {author} {\bibfnamefont {M.}~\bibnamefont {Ernoult}},
  \bibinfo {author} {\bibfnamefont {J.}~\bibnamefont {Kendall}}, \ and\
  \bibinfo {author} {\bibfnamefont {S.}~\bibnamefont {Kumar}},\ }\href@noop {}
  {\enquote {\bibinfo {title} {Energy-based learning algorithms for analog
  computing: a comparative study},}\ } (\bibinfo {year} {2023}),\ \Eprint
  {http://arxiv.org/abs/2312.15103} {arXiv:2312.15103 [cs.LG]} \BibitemShut
  {NoStop}%
\bibitem [{\citenamefont {Kendall}\ \emph {et~al.}(2020)\citenamefont
  {Kendall}, \citenamefont {Pantone}, \citenamefont {Manickavasagam},
  \citenamefont {Bengio},\ and\ \citenamefont
  {Scellier}}]{kendall2020training}%
  \BibitemOpen
  \bibfield  {author} {\bibinfo {author} {\bibfnamefont {J.}~\bibnamefont
  {Kendall}}, \bibinfo {author} {\bibfnamefont {R.}~\bibnamefont {Pantone}},
  \bibinfo {author} {\bibfnamefont {K.}~\bibnamefont {Manickavasagam}},
  \bibinfo {author} {\bibfnamefont {Y.}~\bibnamefont {Bengio}}, \ and\ \bibinfo
  {author} {\bibfnamefont {B.}~\bibnamefont {Scellier}},\ }\href@noop {}
  {\bibfield  {journal} {\bibinfo  {journal} {arXiv preprint arXiv:2006.01981}\
  } (\bibinfo {year} {2020})}\BibitemShut {NoStop}%
\bibitem [{\citenamefont {Martin}\ \emph {et~al.}(2021)\citenamefont {Martin},
  \citenamefont {Ernoult}, \citenamefont {Laydevant}, \citenamefont {Li},
  \citenamefont {Querlioz}, \citenamefont {Petrisor},\ and\ \citenamefont
  {Grollier}}]{martin2021eqspike}%
  \BibitemOpen
  \bibfield  {author} {\bibinfo {author} {\bibfnamefont {E.}~\bibnamefont
  {Martin}}, \bibinfo {author} {\bibfnamefont {M.}~\bibnamefont {Ernoult}},
  \bibinfo {author} {\bibfnamefont {J.}~\bibnamefont {Laydevant}}, \bibinfo
  {author} {\bibfnamefont {S.}~\bibnamefont {Li}}, \bibinfo {author}
  {\bibfnamefont {D.}~\bibnamefont {Querlioz}}, \bibinfo {author}
  {\bibfnamefont {T.}~\bibnamefont {Petrisor}}, \ and\ \bibinfo {author}
  {\bibfnamefont {J.}~\bibnamefont {Grollier}},\ }\href@noop {} {\bibfield
  {journal} {\bibinfo  {journal} {Iscience}\ }\textbf {\bibinfo {volume} {24}}
  (\bibinfo {year} {2021})}\BibitemShut {NoStop}%
\bibitem [{\citenamefont {Stern}\ \emph {et~al.}(2023)\citenamefont {Stern},
  \citenamefont {Liu},\ and\ \citenamefont
  {Balasubramanian}}]{stern2023physical}%
  \BibitemOpen
  \bibfield  {author} {\bibinfo {author} {\bibfnamefont {M.}~\bibnamefont
  {Stern}}, \bibinfo {author} {\bibfnamefont {A.~J.}\ \bibnamefont {Liu}}, \
  and\ \bibinfo {author} {\bibfnamefont {V.}~\bibnamefont {Balasubramanian}},\
  }\href@noop {} {\enquote {\bibinfo {title} {The physical effects of
  learning},}\ } (\bibinfo {year} {2023}),\ \Eprint
  {http://arxiv.org/abs/2306.12928} {arXiv:2306.12928 [cond-mat.dis-nn]}
  \BibitemShut {NoStop}%
\bibitem [{\citenamefont {Ernoult}\ \emph {et~al.}(2020)\citenamefont
  {Ernoult}, \citenamefont {Grollier}, \citenamefont {Querlioz}, \citenamefont
  {Bengio},\ and\ \citenamefont {Scellier}}]{ernoult2020equilibrium}%
  \BibitemOpen
  \bibfield  {author} {\bibinfo {author} {\bibfnamefont {M.}~\bibnamefont
  {Ernoult}}, \bibinfo {author} {\bibfnamefont {J.}~\bibnamefont {Grollier}},
  \bibinfo {author} {\bibfnamefont {D.}~\bibnamefont {Querlioz}}, \bibinfo
  {author} {\bibfnamefont {Y.}~\bibnamefont {Bengio}}, \ and\ \bibinfo {author}
  {\bibfnamefont {B.}~\bibnamefont {Scellier}},\ }\href@noop {} {\bibfield
  {journal} {\bibinfo  {journal} {arXiv preprint arXiv:2005.04168}\ } (\bibinfo
  {year} {2020})}\BibitemShut {NoStop}%
\bibitem [{\citenamefont {Scellier}\ \emph {et~al.}(2022)\citenamefont
  {Scellier}, \citenamefont {Mishra}, \citenamefont {Bengio},\ and\
  \citenamefont {Ollivier}}]{scellier2022agnostic}%
  \BibitemOpen
  \bibfield  {author} {\bibinfo {author} {\bibfnamefont {B.}~\bibnamefont
  {Scellier}}, \bibinfo {author} {\bibfnamefont {S.}~\bibnamefont {Mishra}},
  \bibinfo {author} {\bibfnamefont {Y.}~\bibnamefont {Bengio}}, \ and\ \bibinfo
  {author} {\bibfnamefont {Y.}~\bibnamefont {Ollivier}},\ }\href@noop {}
  {\enquote {\bibinfo {title} {Agnostic physics-driven deep learning},}\ }
  (\bibinfo {year} {2022}),\ \Eprint {http://arxiv.org/abs/2205.15021}
  {arXiv:2205.15021 [cs.LG]} \BibitemShut {NoStop}%
\bibitem [{\citenamefont {Falk}\ \emph
  {et~al.}(2023{\natexlab{b}})\citenamefont {Falk}, \citenamefont {Strupp},
  \citenamefont {Scellier},\ and\ \citenamefont
  {Murugan}}]{falk2023contrastive}%
  \BibitemOpen
  \bibfield  {author} {\bibinfo {author} {\bibfnamefont {M.}~\bibnamefont
  {Falk}}, \bibinfo {author} {\bibfnamefont {A.}~\bibnamefont {Strupp}},
  \bibinfo {author} {\bibfnamefont {B.}~\bibnamefont {Scellier}}, \ and\
  \bibinfo {author} {\bibfnamefont {A.}~\bibnamefont {Murugan}},\ }\href@noop
  {} {\enquote {\bibinfo {title} {Contrastive learning through non-equilibrium
  memory},}\ } (\bibinfo {year} {2023}{\natexlab{b}}),\ \Eprint
  {http://arxiv.org/abs/2312.17723} {arXiv:2312.17723 [cond-mat.dis-nn]}
  \BibitemShut {NoStop}%
\bibitem [{\citenamefont {Dillavou}\ \emph {et~al.}(2022)\citenamefont
  {Dillavou}, \citenamefont {Stern}, \citenamefont {Liu},\ and\ \citenamefont
  {Durian}}]{dillavou2022demonstration}%
  \BibitemOpen
  \bibfield  {author} {\bibinfo {author} {\bibfnamefont {S.}~\bibnamefont
  {Dillavou}}, \bibinfo {author} {\bibfnamefont {M.}~\bibnamefont {Stern}},
  \bibinfo {author} {\bibfnamefont {A.~J.}\ \bibnamefont {Liu}}, \ and\
  \bibinfo {author} {\bibfnamefont {D.~J.}\ \bibnamefont {Durian}},\
  }\href@noop {} {\bibfield  {journal} {\bibinfo  {journal} {Physical Review
  Applied}\ }\textbf {\bibinfo {volume} {18}},\ \bibinfo {pages} {014040}
  (\bibinfo {year} {2022})}\BibitemShut {NoStop}%
\bibitem [{\citenamefont {Wycoff}\ \emph {et~al.}(2022)\citenamefont {Wycoff},
  \citenamefont {Dillavou}, \citenamefont {Stern}, \citenamefont {Liu},\ and\
  \citenamefont {Durian}}]{wycoff2022desynchronous}%
  \BibitemOpen
  \bibfield  {author} {\bibinfo {author} {\bibfnamefont {J.~F.}\ \bibnamefont
  {Wycoff}}, \bibinfo {author} {\bibfnamefont {S.}~\bibnamefont {Dillavou}},
  \bibinfo {author} {\bibfnamefont {M.}~\bibnamefont {Stern}}, \bibinfo
  {author} {\bibfnamefont {A.~J.}\ \bibnamefont {Liu}}, \ and\ \bibinfo
  {author} {\bibfnamefont {D.~J.}\ \bibnamefont {Durian}},\ }\href@noop {}
  {\bibfield  {journal} {\bibinfo  {journal} {The Journal of Chemical Physics}\
  }\textbf {\bibinfo {volume} {156}} (\bibinfo {year} {2022})}\BibitemShut
  {NoStop}%
\bibitem [{\citenamefont {Stern}\ \emph {et~al.}(2022)\citenamefont {Stern},
  \citenamefont {Dillavou}, \citenamefont {Miskin}, \citenamefont {Durian},\
  and\ \citenamefont {Liu}}]{stern2022physical}%
  \BibitemOpen
  \bibfield  {author} {\bibinfo {author} {\bibfnamefont {M.}~\bibnamefont
  {Stern}}, \bibinfo {author} {\bibfnamefont {S.}~\bibnamefont {Dillavou}},
  \bibinfo {author} {\bibfnamefont {M.~Z.}\ \bibnamefont {Miskin}}, \bibinfo
  {author} {\bibfnamefont {D.~J.}\ \bibnamefont {Durian}}, \ and\ \bibinfo
  {author} {\bibfnamefont {A.~J.}\ \bibnamefont {Liu}},\ }\href@noop {}
  {\bibfield  {journal} {\bibinfo  {journal} {Physical Review Research}\
  }\textbf {\bibinfo {volume} {4}},\ \bibinfo {pages} {L022037} (\bibinfo
  {year} {2022})}\BibitemShut {NoStop}%
\bibitem [{\citenamefont {Dillavou}\ \emph {et~al.}(2023)\citenamefont
  {Dillavou}, \citenamefont {Beyer}, \citenamefont {Stern}, \citenamefont
  {Miskin}, \citenamefont {Liu},\ and\ \citenamefont
  {Durian}}]{dillavou2023machine}%
  \BibitemOpen
  \bibfield  {author} {\bibinfo {author} {\bibfnamefont {S.}~\bibnamefont
  {Dillavou}}, \bibinfo {author} {\bibfnamefont {B.~D.}\ \bibnamefont {Beyer}},
  \bibinfo {author} {\bibfnamefont {M.}~\bibnamefont {Stern}}, \bibinfo
  {author} {\bibfnamefont {M.~Z.}\ \bibnamefont {Miskin}}, \bibinfo {author}
  {\bibfnamefont {A.~J.}\ \bibnamefont {Liu}}, \ and\ \bibinfo {author}
  {\bibfnamefont {D.~J.}\ \bibnamefont {Durian}},\ }\href@noop {} {\bibfield
  {journal} {\bibinfo  {journal} {arXiv preprint arXiv:2311.00537}\ } (\bibinfo
  {year} {2023})}\BibitemShut {NoStop}%
\bibitem [{\citenamefont {Laydevant}\ \emph {et~al.}(2023)\citenamefont
  {Laydevant}, \citenamefont {Markovic},\ and\ \citenamefont
  {Grollier}}]{laydevant2023training}%
  \BibitemOpen
  \bibfield  {author} {\bibinfo {author} {\bibfnamefont {J.}~\bibnamefont
  {Laydevant}}, \bibinfo {author} {\bibfnamefont {D.}~\bibnamefont {Markovic}},
  \ and\ \bibinfo {author} {\bibfnamefont {J.}~\bibnamefont {Grollier}},\
  }\href@noop {} {\enquote {\bibinfo {title} {Training an ising machine with
  equilibrium propagation},}\ } (\bibinfo {year} {2023}),\ \Eprint
  {http://arxiv.org/abs/2305.18321} {arXiv:2305.18321 [cs.NE]} \BibitemShut
  {NoStop}%
\bibitem [{\citenamefont {Yi}\ \emph {et~al.}(2023)\citenamefont {Yi},
  \citenamefont {Kendall}, \citenamefont {Williams},\ and\ \citenamefont
  {Kumar}}]{yi2023activity}%
  \BibitemOpen
  \bibfield  {author} {\bibinfo {author} {\bibfnamefont {S.-i.}\ \bibnamefont
  {Yi}}, \bibinfo {author} {\bibfnamefont {J.~D.}\ \bibnamefont {Kendall}},
  \bibinfo {author} {\bibfnamefont {R.~S.}\ \bibnamefont {Williams}}, \ and\
  \bibinfo {author} {\bibfnamefont {S.}~\bibnamefont {Kumar}},\ }\href@noop {}
  {\bibfield  {journal} {\bibinfo  {journal} {Nature Electronics}\ }\textbf
  {\bibinfo {volume} {6}},\ \bibinfo {pages} {45} (\bibinfo {year}
  {2023})}\BibitemShut {NoStop}%
\bibitem [{\citenamefont {Oh}\ \emph {et~al.}(2023)\citenamefont {Oh},
  \citenamefont {An}, \citenamefont {Cho}, \citenamefont {Yoon},\ and\
  \citenamefont {Min}}]{oh2023memristor}%
  \BibitemOpen
  \bibfield  {author} {\bibinfo {author} {\bibfnamefont {S.}~\bibnamefont
  {Oh}}, \bibinfo {author} {\bibfnamefont {J.}~\bibnamefont {An}}, \bibinfo
  {author} {\bibfnamefont {S.}~\bibnamefont {Cho}}, \bibinfo {author}
  {\bibfnamefont {R.}~\bibnamefont {Yoon}}, \ and\ \bibinfo {author}
  {\bibfnamefont {K.-S.}\ \bibnamefont {Min}},\ }\href@noop {} {\bibfield
  {journal} {\bibinfo  {journal} {Micromachines}\ }\textbf {\bibinfo {volume}
  {14}},\ \bibinfo {pages} {1367} (\bibinfo {year} {2023})}\BibitemShut
  {NoStop}%
\bibitem [{\citenamefont {Laborieux}\ \emph {et~al.}(2021)\citenamefont
  {Laborieux}, \citenamefont {Ernoult}, \citenamefont {Scellier}, \citenamefont
  {Bengio}, \citenamefont {Grollier},\ and\ \citenamefont
  {Querlioz}}]{laborieux2021scaling}%
  \BibitemOpen
  \bibfield  {author} {\bibinfo {author} {\bibfnamefont {A.}~\bibnamefont
  {Laborieux}}, \bibinfo {author} {\bibfnamefont {M.}~\bibnamefont {Ernoult}},
  \bibinfo {author} {\bibfnamefont {B.}~\bibnamefont {Scellier}}, \bibinfo
  {author} {\bibfnamefont {Y.}~\bibnamefont {Bengio}}, \bibinfo {author}
  {\bibfnamefont {J.}~\bibnamefont {Grollier}}, \ and\ \bibinfo {author}
  {\bibfnamefont {D.}~\bibnamefont {Querlioz}},\ }\href@noop {} {\bibfield
  {journal} {\bibinfo  {journal} {Frontiers in neuroscience}\ }\textbf
  {\bibinfo {volume} {15}},\ \bibinfo {pages} {633674} (\bibinfo {year}
  {2021})}\BibitemShut {NoStop}%
\bibitem [{\citenamefont {Edwards}\ and\ \citenamefont
  {Anderson}(1976)}]{S_F_Edwards_1976}%
  \BibitemOpen
  \bibfield  {author} {\bibinfo {author} {\bibfnamefont {S.~F.}\ \bibnamefont
  {Edwards}}\ and\ \bibinfo {author} {\bibfnamefont {P.~W.}\ \bibnamefont
  {Anderson}},\ }\href {\doibase 10.1088/0305-4608/6/10/022} {\bibfield
  {journal} {\bibinfo  {journal} {Journal of Physics F: Metal Physics}\
  }\textbf {\bibinfo {volume} {6}},\ \bibinfo {pages} {1927} (\bibinfo {year}
  {1976})}\BibitemShut {NoStop}%
\bibitem [{\citenamefont {Alpaydin}\ and\ \citenamefont
  {Kaynak}(1998)}]{misc_optical_recognition_of_handwritten_digits_80}%
  \BibitemOpen
  \bibfield  {author} {\bibinfo {author} {\bibfnamefont {E.}~\bibnamefont
  {Alpaydin}}\ and\ \bibinfo {author} {\bibfnamefont {C.}~\bibnamefont
  {Kaynak}},\ }\href@noop {} {\enquote {\bibinfo {title} {{Optical Recognition
  of Handwritten Digits}},}\ }\bibinfo {howpublished} {UCI Machine Learning
  Repository} (\bibinfo {year} {1998}),\ \bibinfo {note} {{DOI}:
  https://doi.org/10.24432/C50P49}\BibitemShut {NoStop}%
\bibitem [{\citenamefont {Glorot}\ and\ \citenamefont
  {Bengio}(2010)}]{glorot2010understanding}%
  \BibitemOpen
  \bibfield  {author} {\bibinfo {author} {\bibfnamefont {X.}~\bibnamefont
  {Glorot}}\ and\ \bibinfo {author} {\bibfnamefont {Y.}~\bibnamefont
  {Bengio}},\ }in\ \href@noop {} {\emph {\bibinfo {booktitle} {Proceedings of
  the thirteenth international conference on artificial intelligence and
  statistics}}}\ (\bibinfo {organization} {JMLR Workshop and Conference
  Proceedings},\ \bibinfo {year} {2010})\ pp.\ \bibinfo {pages}
  {249--256}\BibitemShut {NoStop}%
\end{thebibliography}%

    \appendix
    \section{Interpretation of the Cost Function}\label{Appendix:interpretation}

    In Eq.~\eqref{eq:cost_function} we introduced a cost function different from the straightforward distance function in Eq.~\eqref{eq:similarityFunction}. Here we comment further on the problems posed by the distance function. In addition, we point out an interesting connection between the adopted cost function and the overlap of quantum states. 

    From Eq.~\eqref{eq:similarityFunction}, we find that the training reaches its fixed points when 
    \begin{equation}
        \begin{split}
            \sin(\phi_i - \phi_i^\tau) = 0
        \end{split}
    \end{equation}
    holds for the output units. The distance function $D$ reaches it minimum when $\phi_i = \phi_i^T$ and maximum when $\phi_i = \phi_i^T + \pi$. Theoretically, the maxima are unstable and therefore should be avoided. However, practically they can still cause a long stay on some plateaus during training (with $\phi_i = \phi_i^T + \pi$ for some output unit) and seriously increase the time cost for training. In contrast to this behaviour of $D$, the only fixed point for the adopted cost function $C$ is the case when $\phi_i = \phi_i^T$ holds for all the output units. We find that this makes the training much more efficient. 

    Although $C$ was initially chosen simply to provide a strong repulsion from the unstable fixed points (maxima), we can draw an interesting connection to quantum physics. Assume that we have an array of independent two-level quantum systems and prepare the state of the system based on the output units by letting
    \begin{equation}
        |\Phi_i \rangle =  \frac{1}{\sqrt{2}}(|0 \rangle + e^{i \phi_i} |1\rangle )
    \end{equation}
    which is exactly an eigenstate of $\cos \phi_i X + \sin \phi_i Y$ of eigenvalue $+1$ (Here $X = |0\rangle \langle 1 | + |1\rangle \langle 0 |$ and $Y = \i |0\rangle \langle 1 | - \i |1\rangle \langle 0 |$). 

    We prepare another set of two-level systems based on the state of the target in the same way. Then we have two product states, corresponding to the actual output $|S \rangle$ and the target $| T \rangle$
    \begin{equation}
        \begin{split}
            & |S \rangle = \prod_{i \in S_{\rm out}} |\Phi_i \rangle = \frac{1}{2^{|S_{\rm out}|/2}} \prod_{i \in S_{\rm out}} (|0 \rangle + e^{i \phi_i} |1\rangle ) \\
            & | T \rangle = 
            \prod_{i \in S_{\rm out}} 
            |\Phi_i^\tau \rangle = \frac{1}{2^{|S_{\rm out}|/2}} 
            \prod_{i \in S_{\rm out}} (|0 \rangle + e^{i \phi_i^\tau} |1\rangle ).
        \end{split}
    \end{equation}
    Then the task is to minimize the distance between $|S \rangle$ and $|T \rangle$ by maximizing $ \left| \langle T|S \rangle \right| ^2$. This is equivalent to minimizing
    \begin{equation}
        \begin{split}
            \mathcal{L} &= -\log \left| \langle T|S \rangle \right| ^2 \\
            &= -\sum_{i \in S_{out}} \log \left( \frac{1}{4} |1+e^{\i (\phi^\tau_i-\phi_i) }|^2 \right) \\
            &= -\sum_{i \in S_{out}} \log \left( \frac{1}{2} (1+\cos(\phi_i -\phi^\tau_i)) \right)
        \end{split}
    \end{equation}
    which has the form of $C$ expressed in Eq.~\eqref{eq:cost_function}. We note that $ \left| \langle T|S \rangle \right| ^2$ is the probability of measuring the quantum state $|T\rangle$  after preparing the state $| S \rangle$ (if measurements are performed with respect to a basis containing $| T \rangle$ as one of the states). Thus, $\mathcal L$ can also be seen as a categorical cross-entropy $- \sum P_j \ln Q_j$, for the case where $P_1=1$ is the desired probability of outcome ``T'' in such a measurement, and $Q_1 = \left| \langle T|S \rangle \right| ^2$ is the probability of actually observing this outcome.

    \section{Digit recognition performance and benchmarks}
        \label{sec:accuracies}
        We compared the performance of the neuromorphic system to artificial neural networks (ANNs) and linear classifiers which we implemented with tensorflow. The architecture of the models we implemented as well as the number of parameters matched those of the layered structure of our neuromorphic systems. We used a categorical cross-entropy loss function for the ANNs and a mean-squared-error loss function for the linear classifiers and trained both models using Adam optimization for 100 epochs dividing the training data into randomised mini-batches of 10 images. The maximum test accuracy is reported in Tab.~\ref{tab:accuracies} and compared to the performance of the corresponding neuromorphic systems with either layer structure or with all-to-all connectivity.

                \begin{table*}[htbp]
                \centering
                \begin{tabular}{ |c|c|c|c|c|c|c|c| }
                     \hline
                     \multicolumn{3}{|c|}{Layered} &
                     \multicolumn{3}{|c|}{All-to-All} &
                     \multicolumn{2}{|c|}{Benchmark} \\
                     \hline
                     \# hidden units & \# parameters & test accuracy & \# hidden units & \# parameters & test accuracy & ANN & linear classifier \\
                     \hline
                     
                     20   & 1540 & $91.9 \, \%$
                     & 11 & 1596 & $93.3 \,\%$
                     & $94.3 \, \%$ & $90.4 \, \%$ \\
                     100 &   7620  & $92.7 \, \%$
                     & 75 & 7725 & $92.7 \, \%$
                     & $95.0 \, \%$ & $90.3 \, \%$ \\
                     200 & 15220 & $94.1 \, \%$
                     & 115 & 15246 & $92.4 \, \%$
                     & $95.0 \, \%$ & $90.7 \, \%$ \\
                     300 & 22820 & $93.7 \, \%$
                     & 148 & 22831 & $92.9 \, \%$
                     & $95.1 \, \%$ & $90.7 \, \%$ \\
                     
                     \hline
                \end{tabular}
                \caption{\textbf{Digit recognition test accuracy and benchmarks.}
                Benchmarks were obtained with networks of the same structure as the layered neuromorphic system. For the linear classifier, we used linear activations and a mean-squared-error cost function; for the artifical neural network, we used a sigmoid activation function in the second layer, a softmax in the last layer and a categorical cross-entropy activation function.
                }
                \label{tab:accuracies}
            \end{table*}

\end{document}